\documentclass[%
 amsmath,amssymb,
 aps,prl,
 showpacs,
 superscriptaddress,
twocolumn
]{revtex4-2}
\UseRawInputEncoding
\usepackage{graphicx}
\usepackage{dcolumn}
\usepackage{bm}
\usepackage{xcolor}

\usepackage{hyperref}

\newcommand{\red}[1]{\textcolor{black}{#1}}

\begin{document}

\preprint{APS/123-QED}

\title{
Irreversible mesoscale fluctuations herald the emergence of dynamical phases
}

\author{Thomas Suchanek}
 \affiliation{Institut f{\"u}r Theoretische Physik, Universit{\"a}t Leipzig, Postfach 100 920, D-04009 Leipzig, GER}
\author{Klaus Kroy}
\affiliation{Institut f{\"u}r Theoretische Physik, Universit{\"a}t Leipzig, Postfach 100 920, D-04009 Leipzig, GER}
\author{Sarah A. M. Loos}
 \email{sl2127@cam.ac.uk}
\affiliation{DAMTP, Centre for Mathematical Sciences, 
University of Cambridge, Wilberforce Road, Cambridge CB3 0WA, UK
}%

\date{\today}

\begin{abstract} 
We study fluctuating field models with spontaneously emerging
dynamical phases. We consider two typical transition scenarios associated with {parity-time} symmetry breaking: oscillatory instabilities and critical exceptional points. 
An analytical investigation of the {low}-noise \red{regime} reveals \red{a drastic}
increase of the mesoscopic entropy production toward the transitions. 
For an illustrative model of two nonreciprocally coupled Cahn-Hilliard fields, we find physical interpretations in terms of actively propelled interfaces and a coupling of modes near the critical exceptional point.
%
\end{abstract}

\maketitle



When Einstein's paper on Brownian motion appeared, he received a critical letter from C. R\"ontgen~\cite{einstein1993collected} who reiterated the historically widespread concern that such ``motion from heat'' would violate the second law.
He failed to understand that Einstein had just made precise the centuries-old notion of heat being but a name 
for the incessant random motion of the molecular constituents of all macroscopic matter~\cite{hooke2019posthumous}. In fact, these thermal fluctuations are a manifestation of conservation, 
not production, of energy and entropy,
\red{according to the fluctuation-dissipation theorem~\cite{Kubo_1966}}. However, this cornerstone of modern equilibrium statistical mechanics 
is lost far from equilibrium, where it becomes a central task to understand whether, when, and why mesoscopic fluctuations 
produce entropy. 
With regard to biological systems, this literally becomes a question of ``life and death.''

A fundamental property of any thermal equilibrium is 
time-reversal symmetry. Its breaking, in turn, \red{is associated with} a production of entropy and {dissipative} dynamics. 
{Notably, on the coarse-grained scale, however, only part of the full entropy production is generally perceptible.}
A versatile {measure} to quantify time-reversal symmetry breaking (TRSB) is \red{then} the log-ratio of probabilities for forward and backward paths \red{of the coarse-grained dynamics}~\cite{lebowitz1999,maes2003time,Seifert2005,Rold_n_2012}. This so-called (informatic) entropy production rate {$\mathcal{S}$} and related measures 
have recently been studied for a variety of systems; from single or few particle models~\cite{Seifert2005}, over nonequilibrium field theories of active matter~\cite{Li_2021,Nardini2017,seara2021,Borthne2020,pruessner2022field}, to experimental studies on multiscale biological systems~\cite{ro2022model,tan2021scale,battle2016broken}. 

In this Letter, we explore how TRSB of mesocale fluctuations in many-body systems informs us about incipient pattern formation.
Specifically, we address {the TRSB associated with the emergence of} \textit{dynamical} mesophases, such as persistent traveling {or oscillating patterns}~\cite{BESTEHORN1989,bestehorn19892,Ghosh2021,Ramaswamy2000,Mayank17,frey2018protein,huang2003dynamic,Amari1977,Wang15,Bhattacharya20,Welch2002,Takada22,demarchi2023enzyme}.
Such states are paradigmatic examples of dissipative structures maintained by permanent dissipative energy currents~\cite{Goldbeter18,Cross93,Tiezzi2008,Belintsev_1983}.
Their spontaneous emergence is an instructive example of how a hidden nonequilibrium condition and its entropy production may reveal themselves mesoscopically. For example, for the \red{so-called} Brusselator model, it was recently shown that $\mathcal{S}$ may display a significant increase across the static-dynamic phase transitions to its oscillating phase~\cite{seara2021}.

\red{In the following, we consider} a broad class of nonequilibrium field models with conserved dynamics, and focus on two most common static-dynamic transition scenarios:  
oscillatory instabilities~\cite{Cross93,cross_greenside_2009}\red{, which are field-theoretical manifestations of a Hopf bifurcation,} and critical exceptional points (CEP)~\cite{hanai2020critical,Fruchart2021}, 
\red{which arise by coalescence of a Goldstone and a critical mode}.
Both these very dissimilar scenarios can be addressed within the framework of non-Hermitian field theories~\cite{suchanek2023connection}. As was described only recently, CEPs generically occur in many-body systems with 
{nonreciprocal} interactions~\cite{You_2020,Saha_2020,frohoff2021suppression,Fruchart2021,frohoff2023nonreciprocal,Mandal22}, i.e., interactions that violate the action-reaction principle~\cite{loos2020irreversibility}. 
For both scenarios, we show that the approach from the static toward the dynamic phase is accompanied by a surge in entropy production that scales like the susceptibility. 
Further, we discover general connections between the parity-time ($\mathcal{PT}$) symmetry breaking that generically occurs at CEPs\red{~\cite{el2018non,krasnok2021paritytime,Fruchart2021}}, on the one hand, and the emergence of irreversible, i.e., TRSB fluctuations, on the other hand. Below,  and \red{more comprehensively} in a companion article~\cite{suchanek2023entropy}, we illustrate our general findings with a model of two nonreciprocally coupled Cahn-Hilliard field equations~\cite{You_2020,Saha_2020,frohoff2021suppression,frohoff2023nonreciprocal}.


\textit{Field theory and symmetry breaking.---}
We study hydrodynamic models of the following structure
\begin{equation}\label{equ:mod}
    \dot  \phi_i=-\nabla\cdot \boldsymbol{J}_i,
    ~~\boldsymbol{J}_i=-\nabla\mu_i+\sqrt{2\epsilon}\boldsymbol{\Lambda}_i\,,
\end{equation}
\red{with $N$ scalar field components $\phi_i(\boldsymbol{r},t)$, $i=1,\dots,N$, representing conserved order parameters, such as the species number densities, of an active many-body system and their currents $\boldsymbol{J}_i\red{(\boldsymbol{r},t)}$
~\cite{Mandal22,knevzevic2022}.}  
 The Gaussian \red{space-time} white noise term $\sqrt{2\epsilon}{\boldsymbol{\Lambda}}_i\red{(\boldsymbol{r},t)}$ is constructed such that, in the equilibrium case, \red{where
$\mu_i[\phi]$ derives from a free energy functional $\mathcal{F[{\phi}]}$, i.e.
$\mu_i={\delta \mathcal{F}}/{\delta{\phi_i}}$,} the resulting statistical field theory would obey a fluctuation-dissipation relation~\cite{Kubo_1966}, with $\epsilon$ denoting the noise intensity. However, for the case of a nonequilibrium deterministic current 
\red{$\boldsymbol{J}_i^\mathrm{d}[\phi]\equiv -\nabla {\mu_i}[\phi]$}, which is of interest here, the chemical potential ${\mu}_i$ cannot be represented as a gradient.
\red{To exclude externally driven systems, we further assume that Eq.~\eqref{equ:mod} is {invariant} with respect to parity inversion, $\mathcal{P}: \boldsymbol{r}\mapsto-\boldsymbol{r}$.}

A first useful insight is that, by construction, the {spontaneous} emergence of phases with traveling patterns \red{in models of the type of Eq.~\eqref{equ:mod}} is always accompanied by a breaking of $\mathcal{PT}$ symmetry. This can be seen as follows. 
In a phase with traveling patterns, the zero-noise limit ($\epsilon \to 0$) of Eq.~\eqref{equ:mod} has solutions of the form $\phi_i^*(\boldsymbol{r},t)\equiv\varphi_i(\boldsymbol{r}-\boldsymbol{v}t)$. Then, \red{the $\mathcal{P}$ invariance of Eq.~\eqref{equ:mod}}
implies that $ \varphi_i'(\boldsymbol{r}+\boldsymbol{v}t)$, with $\varphi_i'(\boldsymbol{x})=\mathcal{P}\varphi_i(\boldsymbol{x})\red{=\varphi_i(-\boldsymbol{x})}$ is also a solution; which can as well be expressed as $\varphi_i'(\boldsymbol{r}+\boldsymbol{v}t)=\mathcal{T}\varphi_i'(\boldsymbol{r}-\boldsymbol{v}t)$, with $\mathcal{T}$ the time-inversion operator.
Therefore, the $\mathcal{PT}$ operation applied to any given traveling pattern solution of Eq.~\eqref{equ:mod} yields another solution. Now, it is clear that, on the one hand, a parity symmetric pattern (i.e., $\varphi'=\varphi$) can occur only for $\boldsymbol{v}=0$, and that, on the other hand,
spontaneously emerging dynamical solutions ${\varphi}$ with $\boldsymbol{v}\neq0$
automatically cease to be $\mathcal{PT}$ eigenfunctions.   
Note that the emergence of $\mathcal{PT}$-broken dynamical phases is not specific to field models of \red{the type of Eq.~\eqref{equ:mod}} but observed in a much wider context, comprising polar swarm models~\cite{Borthne2020,Fruchart2021}, directional solidification~\cite{Coullet89}, or driven interfaces~\cite {Cummins93,Pan94}.

\textit{Irreversibility.---}
To study irreversibility, we employ a framework~\cite{Li_2021,Nardini2017} that defines the entropy production ${s}[\phi;0,T]$ along a trajectory $\{\phi_{t\in [0,T]}\}$ as the log ratio of forward and backward path probabilities (see Ref.~\cite{suchanek2023connection} for details).
The average rate of entropy production,
$  {{\mathcal{S}}}= \lim_{h\to 0} \langle{s}[\phi;t,t+h]/{h}\rangle,
$
serves as a measure of the breaking of detailed balance and of time-reversal symmetry, at time $t$, \red{where} $\langle .\rangle$ denotes the noise
average.
\red{A main result of Ref.~\cite{Nardini2017} was that it can, in the steady state, be expressed as the volume integral}
$\mathcal{S}=-\epsilon^{-1}\sum_i\int_V\mathrm{d}\boldsymbol{r}\langle\dot{\phi}_i\mu_i\rangle$\red{, which is understood to be UV-regularized}.
\red{By employing It\^o calculus of functionals~\cite{Itofunc}, we derive, as a central result of the companion paper~\cite{suchanek2023connection}, the more explicit form}
\begin{align}\label{equ:EPRhy}
\mathcal{S}=&\int_V\!\mathrm{d}\mathbf{r}
\frac{\sum_i\langle \vert {\boldsymbol{J}}_i^\mathrm{d}\vert^2 \rangle}{\epsilon}+\!
\int_V\!\mathrm{d}\mathbf{r}\sum_i\left\langle\frac{\delta}{\delta{\phi_i}}\nabla\cdot \boldsymbol{J}_i^\mathrm{d} \right\rangle\,,  
\end{align} 
\red{
It yields $\mathcal{S}$ based on the single time probability distribution of  $\phi$ alone, which is particularly useful to study phase transitions.
}


\begin{figure}[t]
\centering
\includegraphics[width=.48\textwidth]{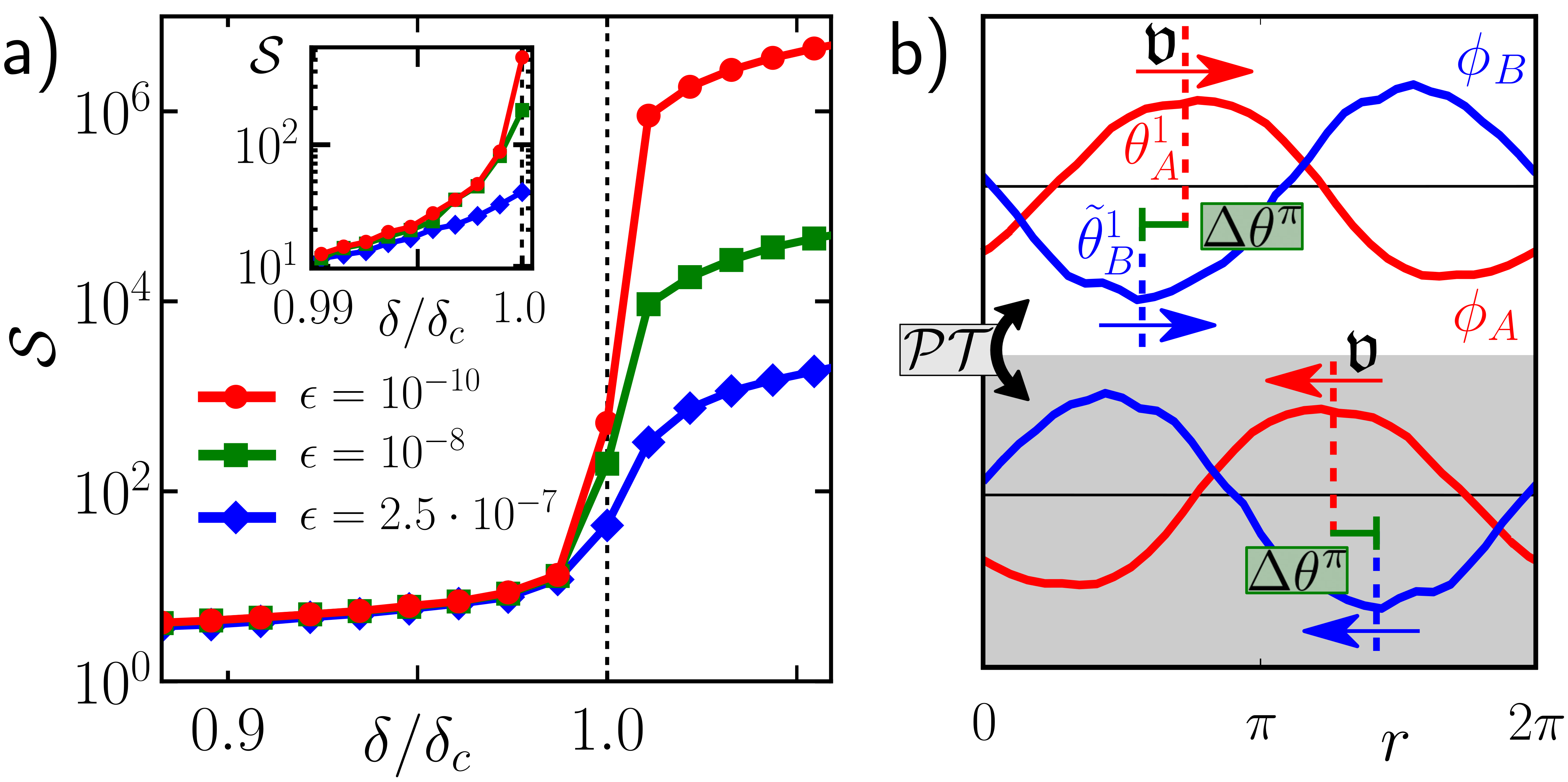}

\caption{\label{fig:pd} 
Transition to a dynamical phase via a CEP.
(a) Entropy production rate near the transition point $\delta_c$ from a static demixed to a traveling-wave phase in model \red{Eq.~\eqref{equ:1}} ($\alpha=-0.07,\gamma=0.015,\kappa=0.01,\beta=0.05$), \red{see Ref.~\cite{suchanek2023entropy} for simulation details. The inset shows a magnification}. (b) Traveling-wave solution of Eq.~\eqref{equ:1} and its image under a parity transformation $\mathcal{P}$ (which yields an independent solution). The characteristic phase shift $\Delta \theta^\pi$ is always aligned with the propagation velocity $\mathfrak{v}$ (indicated by arrows).
}
\end{figure}

We \red{now} consider the linear stability of the zero-noise solutions $\phi^*$ 
of our dynamical system, i.e., the eigensystem of the Jacobian $(\mathcal{J}_{\phi*})_{ij} =-\delta (\nabla \cdot \boldsymbol{J}^\mathrm{d}_i)/\delta \phi_j^*$, represented in a Fourier basis. A mode (eigenvector) becomes unstable when the real part of its eigenvalues vanishes.
Due to the parity symmetry of Eq.~\eqref{equ:mod}, $\mathcal{J}_{\phi*}$ is real. 
Crucially, we allow $\mathcal{J}_{\phi*}$ to be \textit{non-Hermitian}, thereby capturing a wide variety of nonequilibrium conditions~\cite{Fruchart2021,Ashida2020}.
A common route from a (mesoscopically) static state to a dynamical state is via oscillatory instabilities. They occur through a pair of complex conjugated eigenmodes of $\mathcal{J}_{\phi*}$ that become unstable and whose eigenvalues (here denoted by $\lambda_{\pm}$) have non-vanishing imaginary parts. 
In contrast, the alternative scenario of a CEP is not primarily characterized by the properties of the eigenvalues of $\mathcal{J}_{\phi*}$, but by the fact that the modes that loose stability in the course of the transition \textit{align}.
In the following, we assume that a transition involving only two modes $\hat{e}_0$, $\hat{e}_1$ with eigenvalues  $\lambda_0$ and $\lambda_1$.
\red{If one of them is a \red{Goldstone} mode of a broken continuous symmetry (such as space translation invariance or $O(N)$ symmetry~\cite{zelle2023universal}) so that $\lambda_0=0$ for $\hat{e}_0$, the CEP leads into a {dynamical} phase, which dynamically restores the continuous symmetry associated with this Goldstone mode~\cite{Fruchart2021,suchanek2023connection}. }

Using Eq.~\eqref{equ:EPRhy}, we deduce general characteristics of the entropy production rate $\mathcal{S}$ in the vicinity of the transitions. \red{Thereby, the limit
$\mathcal{S}^*\equiv \lim_{\epsilon\to 0}\mathcal{S} 
$ is particularly informative, revealing the leading order contribution to TRSB in $\epsilon$ which dominates the} \red{entire low noise regime and has universal (model-independent) character.} 
Our first general finding is that 
in static phases with non-Hermitian dynamics, $\mathcal{S}^*$ {is generally of order} $\epsilon^0$, i.e., \red{they exhibit} TRSB even \red{for arbitrarily low noise intensity.} Near oscillatory instabilities of monochromatic modes of wavelength $2\pi/\vert \boldsymbol{q}^k\vert$, {$\mathcal{S}^*$} behaves like
\begin{align}\label{equ:scalingEPRatOI}
&
    \mathcal{S}^*\sim \vert \boldsymbol{q}^k\vert^{-2}\vert\mathrm{Im}\lambda_{\pm}\vert^2\chi\,,
    ~\text{as}~
\mathrm{Re}\lambda_{\pm}\rightarrow0\,,
\end{align}
where \red{$\chi \equiv\lim_{\epsilon\rightarrow 0}\epsilon^{-1}\sum_i\int_V\mathrm{d}\boldsymbol{r}\langle \vert\phi_i-\phi_i^*\vert^2\rangle $} is the system's \red{static} susceptibility, which scales like $\vert\mathrm{Re}\lambda_{\pm}\vert^{-1}$, close to the transition.
The proof and the general expression is presented in Ref.~\cite{suchanek2023connection}. 
Similarly, for the CEP, we find
\begin{align}\label{equ:scalingEPRatCEP2}
&
\red{\mathcal{S}^*\propto \chi\propto\lambda_1^{-1}} \,,
~\text{as}~
\lambda_1\rightarrow 0\,.
\end{align}
Here, the most notable insight is that, in any case, we can identify  a  component $\boldsymbol{J}^\mathrm{d}_0\equiv\hat{P}_0\boldsymbol{J}^\mathrm{d}$, of the deterministic current \red{along} $\hat{e}_0$ as the primary source of irreversibility~\footnote{see Ref.~\cite{suchanek2023connection} for the precise definition of the projector $\hat{P}_0$}, so that
$
\mathcal{S}^*$ is dominated by $\int_V\mathrm{d}\boldsymbol{r}\langle\vert \boldsymbol{J}^\mathrm{d}_0\vert^2\rangle\,/\epsilon.
$
This current is generated by
a one-way coupling from damped modes to the Goldstone mode, which entails a giant noise amplification in the latter~\cite{suchanek2023connection}.
From Eqs.~(\ref{equ:scalingEPRatOI}),\,(\ref{equ:scalingEPRatCEP2}), we conclude that, for both types of transitions, $\mathcal{S}$ is determined by the inverse of the real part of the eigenvalue that becomes unstable across the transition. 
Thus, the static phases exhibit, close to the transitions, massively growing (and for $\epsilon \to 0$ \red{diverging}) irreversibility, despite their lack of systematic 
transport, and despite their seemingly equilibrium-like character.
\red{In contrast, across conventional critical points (and generally transitions that are accompanied by the sign change of a single real eigenvalue), $\mathcal{S}$ remains regular, despite diverging $\chi$~\cite{suchanek2023connection}.}

Within the {dynamical} phase itself, \red{the small-noise expansion~\cite{gardiner2009} of}
Eq.~\eqref{equ:EPRhy} \red{with respect to $\phi^*$ yields}~\cite{suchanek2023connection}
\red{
\begin{align}\label{equ:EPRinDP}
    {\mathcal{S}}=
    \epsilon^{-1}\sum_i
    \int_V\mathrm{d}\boldsymbol{r}
 \big\vert \boldsymbol{J}_i^\mathrm{d}(\phi^*)\big\vert^2 
 +\mathcal{O}(\epsilon^0).
\end{align}
Since the deterministic current 
is not sensitive to $\epsilon$, ${\mathcal{S}}\sim \epsilon^{-1}$ to leading order.}
If the {dynamical} phase admits a
traveling pattern $\phi_i^*(\boldsymbol{r},t)\equiv\varphi_i(\boldsymbol{r}-\boldsymbol{v}t)$, 
\red{Eq.~\eqref{equ:EPRinDP} takes the more explicit  form}
$ \mathcal{S}= \epsilon^{-1}\vert \boldsymbol{v}\vert^2\sum_i
    \int_V\mathrm{d}\mathbf{r}
 \vert \varphi_i\vert^2 +\mathcal{O}(\epsilon^0)$,
\red{revealing that $\mathcal{S}$ originates from the hydrodynamic mesoscopic mass fluxes \red{$\boldsymbol{v}\vert\varphi_i\vert$}.}


\textit{Illustrative Example.---}
To illustrate our general findings, we consider a concrete model of \red{the type of Eq.~\eqref{equ:mod}}, namely a stochastic version of the nonreciprocal Cahn-Hilliard model~\cite{You_2020,Saha_2020,frohoff2021suppression,frohoff2023nonreciprocal}. It is simple enough to be analytically traceable, while still exhibiting a \red{dynamical phase, which is accesible via a CEP and an oscillatory instability.}
The dynamical equations for the two-component field ${\phi}=(\phi_A,\phi_B)^T$ read 
\begin{equation}\label{equ:1}
\begin{aligned}
\!\dot \phi_A &= \nabla [ (\alpha\!+\! \phi_A^2 \!-\!\gamma\nabla^2)\nabla \phi_A +\!(\kappa\!-\!\delta)\nabla\phi_B  
+ \!\sqrt{ 2\epsilon}\Lambda_A  ]  \\
\!\dot \phi_B &= \nabla [\beta\nabla \phi_B +(\kappa\!+\!\delta)\nabla\phi_A  
+\sqrt{ 2\epsilon}\Lambda_B  ].
\end{aligned}
\end{equation}
The nonreciprocal coupling $\delta$ between $\phi_A$ and $\phi_B$ ensures that the equations cannot be derived from a scalar potential and represent a non-Hermitian, nonequilibrium model. We study the dynamics on the one-dimensional domain $[0,2\pi]$ with periodic boundary conditions\red{, where the emergence of a dynamical phase itself, and the general features of the phase diagram, do not depend on the system size.}
The noise-free ($\epsilon=0$) case of Eq.~\eqref{equ:1} was shown to exhibit three distinct phases~\cite{You_2020}: a homogeneous phase ($\phi^*_{A,B}=0$) for small negative $\alpha$ and
 two inhomogeneous ``demixed'' phases for large negative $\alpha$.
The approximate solution in terms of the dominant first Fourier mode $\phi^{1,*}_{A,B}(r,t)=\mathcal{A}^{1,*}_{A,B}\cos[r+\theta^{1,*}_{A,B}(t)]$ amounts to a static demixed phase for $\delta<\delta_c=\sqrt{\beta^2+\kappa^2}$, when $\dot \theta^{1,*}_{A,B}(t)=0$, and to a  traveling-wave phase for $\delta>\delta_c$, when $\dot  \theta^{1,*}_{A,B}(t)=\mathfrak{v}\red{=\pm \sqrt{\delta^2-\delta_c^2}}$, \red{with }both signs \red{of the propagation velocity $\mathfrak{v}$} being equally likely. 
The transition from the homogeneous to the traveling-wave state is through an oscillatory instability,
while the (secondary) phase transition from the static-demixed state to the traveling-wave state is via an CEP. 
We simulated Eq.~\eqref{equ:1} using an Euler-Maruyama-algorithm with finite difference gradients, where the domain was discretized by equally spaced mesh points.
For more details about the analytical and numerical treatment,  
we refer to our companion article~\cite{suchanek2023entropy}. Beyond the general Eqs.~(\ref{equ:scalingEPRatOI})--(\ref{equ:EPRinDP}), it confirms that the entropy production scales like $\sim\epsilon^{0}$ in the static (homogeneous and demixed) phases, with singularities {of {$\mathcal{S}^*$}} at the transitions to the traveling-wave phase, and {diverging} {$\mathcal{S}^*$} in the traveling-wave state itself. 
\begin{figure*}[t]

\centering
\includegraphics[width=.99\textwidth]{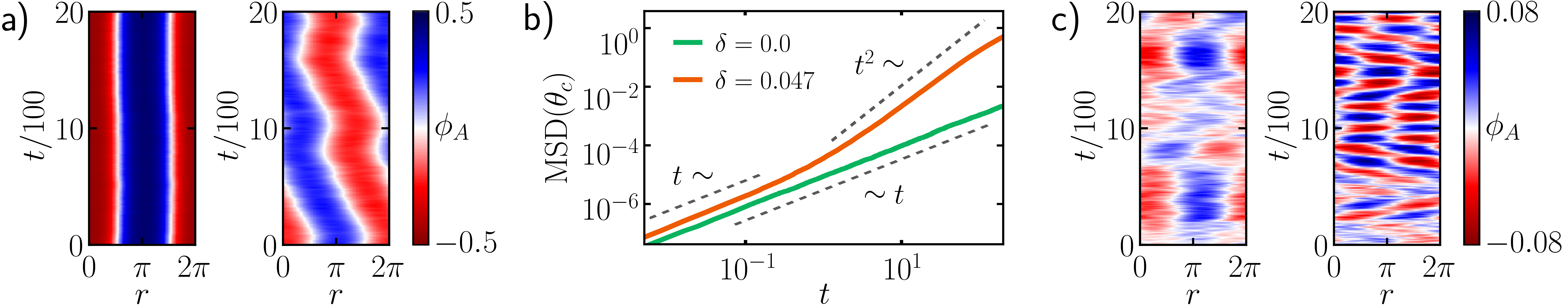}
\caption{Dynamics of the fluctuating Cahn-Hilliard model, Eq.~\eqref{equ:1}, 
for $\epsilon=2.5\times10^{-5}$ in the static phases: (a) Kymographs of the concentration field $\phi_A$ in the static-demixed phase ($\alpha=-0.07$), \textit{left}: for the reciprocal equilibrium case ($\delta=0$); \textit{right}: close to the phase transition ($\delta=0.92\delta_c=0.047$). (b) The corresponding MSD of the ``center-of-mass'' variable  $\theta_c$. The ballistic regime ($\propto t^2$) signals persistent motion of the mean phase and the interfaces. (c) Kymographs of $\phi_A$ in the mixed phase, \textit{left}: for the reciprocal case ($\delta=0, \alpha=-0.013$); \textit{right}: close to the oscillatory instability ($\delta=0.06, \alpha=-0.063$).
\label{fig:MSD}}
\end{figure*}

In the remainder, we exploit the explicit model, \red{Eq.~\eqref{equ:1}}, to provide a physical interpretation of our general findings.
In particular, we reveal which physical mechanism gives rise to the entropy production (and ultimately its divergence), as the transition is approached.

Let us first consider the {dynamical} phase itself. {The sublinear scaling of $\mathcal{S}$ in $\epsilon$, Eq.~\eqref{equ:EPRinDP}, {indicates}} that the {dynamical} phase exhibits a \red{macroscopic arrow of time}. Recalling its definition in terms of path probabilities, a divergent {$\mathcal{S}^*$} means that, upon observation of an (arbitrarily short) realization of the dynamics, one can be $100\%$ sure in which direction time evolves.
Where does this certainty come from? 
The sign of $\mathfrak{v}$ is spontaneously determined by the initial condition and noise, only. Thus, the mere propagation of the wave alone \textit{cannot} introduce an arrow of time. However, closer inspection of the solutions of Eq.~\eqref{equ:1} shows that \red{the propagation velocity} $\mathfrak{v}$ is aligned with a characteristic phase shift $\langle \Delta\theta^{\pi}\rangle \neq 0$, with
$ \Delta \theta^{\pi}\equiv \theta^1_A - (\theta^1_B + \pi)$. The maxima of $\phi_B$ always ``lag behind'' the maxima of $\phi_A$, as shown in Fig.~\ref{fig:pd}(b). The alignment of $\mathfrak{v}$ with $\langle \Delta \theta^{\pi}\rangle$
manifestly breaks $\mathcal{PT}$ symmetry and introduces the macroscopic arrow of time.

Next, we turn to the static phases of Eqs.~\eqref{equ:1} and the unbounded increase of {$\mathcal{S}^*$} toward the static-dynamic transitions predicted by Eqs.~\eqref{equ:scalingEPRatOI},\,\eqref{equ:scalingEPRatCEP2}.
A careful 
decomposition of the phase and amplitude fluctuations of $\phi_{A/B}$ in the static-demixed phase yields an interesting observation: close to the CEP, the fluctuating collective motion of the interfaces between the demixing profiles, represented by $\theta_{A,B}^1(t)$, produces most of the entropy ~\cite{suchanek2023entropy}. As we show here, this entropy-generating interface motion is congruent to the {irreversible motion} of an ``active particle'' or microswimmer and results from systematic activations of the Goldstone mode. Formally, this can be seen as follows.
Within the {low} noise regime, $\theta^1_{A,B}(t)$ can approximately be separated from higher modes, resulting in the closed equations of motion~\cite{suchanek2023entropy}
\begin{align}\label{equ:dyntheta1}
\dot  {\theta}^1(t)=\mathbb{A}\cdot {\theta}^1+ \sqrt{\epsilon/2\pi }\,{\xi},
\end{align}
with ${\theta}^1 = ({\theta}_A^1, \tilde{\theta}_B^{1})^T,$ $\tilde{\theta}_B^{1} =\theta_B^1+\pi$, and 
$\langle\xi_i(t),\xi_j(t')\rangle=(2/\mathcal{A}_{i}^{*})^2 \delta_{i,j}\delta(t-t')$, $i,j \in \{A,B\}$, and
\begin{align}\label{equ:operator}
\mathbb{A} = 
- \begin{pmatrix}
({\kappa^2-\delta^2})/{\beta} & -({\kappa^2-\delta^2})/{\beta}\\
-\beta & \beta\\
\end{pmatrix} \,.
\end{align}
One eigenvalue of the dynamical operator $\mathbb{A}$ is strictly zero, $\lambda_0=0$, reflecting the native continuous translational symmetry of the model \red{in Eq.~\eqref{equ:1}}, which is spontaneously broken in the static-demixed phase (giving rise to a Goldstone mode). The other eigenvalue is $\lambda_1= -(\delta_c^2-\delta^2)/\beta$. 
Now, expressing the phase dynamics~(\ref{equ:dyntheta1}),\,(\ref{equ:operator}) in the ``\textit{center-of-mass}'' frame, by
applying the coordinate transformation $\theta_c\equiv(\Gamma_A\theta_A^1+\Gamma_B\tilde{\theta}_B^1)/\bar{\Gamma}$, 
$\Delta\widetilde{\theta}^\pi\equiv{\sqrt{\Gamma_A\Gamma_B}}{\bar\Gamma^{-1}}\,\Delta\theta^\pi$,
with $\Gamma_{A,B}\equiv \int_{0}^{2\pi}\mathrm{d}{r}\vert\phi^{1,*}_{A,B}({r})\vert^2,$ $\bar{\Gamma}=\Gamma_A+\Gamma_B$,
yields
\begin{align}\label{eq:centerofmass}
\partial_t
 \begin{pmatrix}
\theta_c
\\
{\Delta\widetilde{\theta}^\pi}
\end{pmatrix} 
=
\begin{pmatrix}
0 & 2
\delta \\
0 & \lambda_1
\end{pmatrix}
 \begin{pmatrix}
\theta_c
\\
{\Delta\widetilde{\theta}^\pi}
\end{pmatrix} 
+
\sqrt{2\epsilon/\bar{\Gamma}}\,
\tilde{\xi}\,,
\end{align}
with $\langle \tilde{\xi}_\mu\tilde{\xi}_\nu\rangle=\delta_{\mu \nu}\delta(t-t')$. It reveals that, for $\vert\delta\vert> 0$,  $\theta_c$ performs a persistent random walk ``\red{propelled}'' by the fluctuating phase shift $\Delta \theta^\pi $, in striking analogy to an active particle, with propulsion velocity $\mathfrak{v}_\mathrm{a}\equiv 2\delta\,\Delta\widetilde{\theta}^\pi$ in the \red{Active Ornstein-Uhlenbeck model}~\cite{fodor2016far,Caprini_2019}. %
The persistence time  $t_p=1/|\lambda_1|$ and 
$\langle\vert\mathfrak{v}_\mathrm{a}\vert^2 \rangle=4\delta^2\epsilon /(\bar{\Gamma}|\lambda_1|)$ are controlled by the inverse eigenvalue $\lambda_1^{-1}$ of the dynamical operator $\mathbb{A}$, which vanishes at the transition. This means that the interface between $\phi_A$ and $\phi_B$ fluctuates like the path of a \red{microswimmer}, which is, by itself,
indicative of 
TRSB~\cite{Falasco16}. Indeed, this is visually evident from the exact numerical kymographs in Fig.~\ref{fig:MSD}(b). To obtain a quantitative comparison with the interface dynamics of equilibrium demixing, we compute from Eq.~\eqref{eq:centerofmass} the mean squared displacement (MSD) of $\theta_c$~\footnote{The MSD of $\theta_c$ is readily obtained by applying the results from Eq.~\cite{Nguyen21} to Eq.~\eqref{eq:centerofmass}. We further note that the MSD of the mean phase
	$\theta_\mathrm{m}\equiv (\theta_A^1+\tilde\theta_B^{1})/2$, which may be more accessible from an experimental perspective, exhibits the same ballistic short-time regime and the identical, enhanced long-time diffusion coefficient.
	}, and find
\begin{align}
    \mathrm{MSD}(\theta_c)=
    \begin{cases}
    \frac{\epsilon}{\bar{\Gamma}}\left(t+\frac{\delta^2}{2\lambda_1}t^2\right) \,,
    &t\ll t_p\\
     \frac{\epsilon}{\bar{\Gamma}}\left(1+\frac{\delta^2}{\lambda_1^2}\right)\,t\,,
    &t\gg t_p\,
    \end{cases}.
\end{align}
Clearly, the ``activity'' for $\vert \delta \vert> 0$ is revealed by a ballistic intermediate regime and a strongly enhanced late-time diffusion coefficient in the vicinity of the transition ($\lambda_1\rightarrow 0$).
The predictions of our approximate solution are nicely confirmed by the numerical data presented in Fig.~\ref{fig:MSD}(a) showing the MSD of the interface dynamics obtained directly from Eq.~\eqref{equ:1}. 
The ``active fluctuations'' of 
the interface dynamics \red{in the static demixed phase} represent a concrete manifestation of TRSB on timescales comparable to the persistence time -- in the same manner as it does for an active swimmer (note that here both fields $\phi_{A,B}$ are even under time-reversal~\cite{shankar2018hidden}). As can immediately be gleaned from Eq.~\eqref{eq:centerofmass}, this TRSB yields a considerable fraction $\mathcal{S}^*_{\theta_c}=\bar{\Gamma}\langle\vert \mathfrak{v}_\mathrm{a}\vert^2\rangle\lesssim\mathcal{S}^*$ of the total entropy production. It is generated by the irreversible {mesoscopic} current $\propto\mathfrak{v}_\mathrm{a}$, pointing \red{along} the Goldstone mode and originating from the dissipative coherent motion of both demixing profiles, represented by $\theta_c$.
As the transition to the {dynamical} phase is approached ($\delta\to \delta_c$), $\mathfrak{v}_\mathrm{a}$ and $t_p$ increase unboundedly, resulting in an unbounded increase of {$\mathcal{S}^*$}. 
Importantly, the persistent motion of $\theta_c $ in Eq.~\eqref{eq:centerofmass} is always oriented \textit{toward} the phase shift $\Delta\theta^\pi$ (which has zero mean, $\langle \Delta\theta^\pi \rangle = 0$, contrasting the \red{permanently} $\mathcal{PT}$ broken traveling-wave state), see Fig.~\ref{fig:pd}(b).
We emphasize that such one-way coupling between the damped modes (here $\Delta\theta^\pi$) and a Goldstone mode (here $\theta_c$) that does not vanish at the transition is a hallmark of CEPs and only possible for non-Hermitian dynamical operators~\cite{suchanek2023connection}.
It can be understood as a maximum violation of detailed balance. In the vicintiy of the CEP, this generically leads to a diversion and amplification of fluctuations into the direction of the Goldstone mode, as was recently described in the context of quantum systems~\cite{hanai2020critical}. As we further elaborate for general field models of \red{the type of Eq.~\eqref{equ:1}} in Ref.~\cite{suchanek2023connection}, this very mechanism is deeply connected with the origin of $\mathcal{PT}$ symmetry breaking at the CEP. 

Lastly, also for the divergence of $\mathcal{S}^*$ at the oscillatory instability, the nonreciprocal Cahn-Hilliard model provides a rather intuitive interpretation. Noting that $\vert\mathrm{Im}\lambda_{\pm}\vert$ corresponds to the frequency of the limit cycle, we find that 
$\mathcal{S}$ signals the presence of cyclic currents {in the homogeneous phase}. Their amplitude is entirely due to transient fluctuations, but eventually becomes systematically positive as $\chi$ diverges at the transition. 
In our example, this mechanism creates temporarily stable traveling waves, which are plainly obvious in the graphical representation of our simulation data in Fig.~\ref{fig:MSD}(c).

\textit{Conclusions.---} 
We have studied the irreversible fluctuations
of coarse-grained hydrodynamic field models close to the onset of {dynamical} phases. 
Clearly, the {dynamical} phase itself is a particularly drastic manifestation of TRSB. But our results show that, even before entering \red{it}, the nonequilibrium character of the dynamics reveals itself through \red{transient} TRSB fluctuations around a seemingly equilibrium-like average behavior. This observation could be of interest for a nonequilibrium classification of living matter~\cite{battle2016broken}.
Focusing on two paradigmatic transition scenarios, \red{we uncovered that the fluctuations near $\mathcal{PT}$-breaking phase transitions
not only inflate, as is in equilibrium critical phenomena, but also develop an asymptotically increasing time-reversal asymmetry and associated surging entropy production.}
In the {low} noise regime, {$\mathcal{S}$}  scales precisely as the susceptibility, or equivalently, as the inverse of the eigenvalue of the mode that becomes unstable across the transition. 
Moreover, we have drawn general connections between the $\mathcal{PT}$-symmetry breaking and TRSB of the fluctuations, at a CEP.  We were able to attribute the striking simultaneous presence of two completely different forms of symmetry breaking ($\mathcal{PT}$ and TRSB) to a common origin, an active amplification of thermal noise into irreversible fluctuations.
Our analysis of a model consisting of two nonreciprocally coupled noisy Cahn-Hilliard fields illustrates our general findings and provides instructive physical interpretations 
for the TRSB in terms of a peculiar phenomenology of ``active interface dynamics''. 
In future studies, it would be interesting to reconsider our findings from the perspective of ``information thermodynamics'', as recently pioneered for Turing patterns~\cite{falasco2018information}, \red{and from the perspective of the renormalization group (RG), as  in Refs.~\cite{Paoluzzi22,caballero2020stealth}.}
\red{Further, it would be worthwhile to investigate the spontaneously emerging mass current in the context of thermodynamics uncertainty relations~\cite{Barato15, Gingrich16,niggemann2020field}.}

Independent, consistent results for TRSB in the nonreciprocal Cahn-Hilliard model were reported in Ref.~\cite{alston2023irreversibility}.

\begin{acknowledgments}
We thank 
{\'E}tienne Fodor and Jeremy O'Byrne for valuable comments. 
SL acknowledges funding by the Deutsche Forschungsgemeinschaft (DFG, German Research Foundation) through project 498288081.
TS acknowledges financial
support by the pre-doc award program at Leipzig University. 
SL thanks the Physics Institutes of Leipzig University for their hospitality during several research stays.
\end{acknowledgments}

\bibliography{bib}

\providecommand{\noopsort}[1]{}\providecommand{\singleletter}[1]{#1}%
\begin{thebibliography}{66}%
\makeatletter
\providecommand \@ifxundefined [1]{%
 \@ifx{#1\undefined}
}%
\providecommand \@ifnum [1]{%
 \ifnum #1\expandafter \@firstoftwo
 \else \expandafter \@secondoftwo
 \fi
}%
\providecommand \@ifx [1]{%
 \ifx #1\expandafter \@firstoftwo
 \else \expandafter \@secondoftwo
 \fi
}%
\providecommand \natexlab [1]{#1}%
\providecommand \enquote  [1]{``#1''}%
\providecommand \bibnamefont  [1]{#1}%
\providecommand \bibfnamefont [1]{#1}%
\providecommand \citenamefont [1]{#1}%
\providecommand \href@noop [0]{\@secondoftwo}%
\providecommand \href [0]{\begingroup \@sanitize@url \@href}%
\providecommand \@href[1]{\@@startlink{#1}\@@href}%
\providecommand \@@href[1]{\endgroup#1\@@endlink}%
\providecommand \@sanitize@url [0]{\catcode `\\12\catcode `\$12\catcode
  `\&12\catcode `\#12\catcode `\^12\catcode `\_12\catcode `\%12\relax}%
\providecommand \@@startlink[1]{}%
\providecommand \@@endlink[0]{}%
\providecommand \url  [0]{\begingroup\@sanitize@url \@url }%
\providecommand \@url [1]{\endgroup\@href {#1}{\urlprefix }}%
\providecommand \urlprefix  [0]{URL }%
\providecommand \Eprint [0]{\href }%
\providecommand \doibase [0]{https://doi.org/}%
\providecommand \selectlanguage [0]{\@gobble}%
\providecommand \bibinfo  [0]{\@secondoftwo}%
\providecommand \bibfield  [0]{\@secondoftwo}%
\providecommand \translation [1]{[#1]}%
\providecommand \BibitemOpen [0]{}%
\providecommand \bibitemStop [0]{}%
\providecommand \bibitemNoStop [0]{.\EOS\space}%
\providecommand \EOS [0]{\spacefactor3000\relax}%
\providecommand \BibitemShut  [1]{\csname bibitem#1\endcsname}%
\let\auto@bib@innerbib\@empty
\bibitem [{\citenamefont {Einstein}\ and\ \citenamefont
  {Klein}(1993)}]{einstein1993collected}%
  \BibitemOpen
  \bibfield  {author} {\bibinfo {author} {\bibfnamefont {A.}~\bibnamefont
  {Einstein}}\ and\ \bibinfo {author} {\bibfnamefont {M.}~\bibnamefont
  {Klein}},\ }\bibinfo {title} {The collected papers of {A}lbert {E}instein:
  The {S}wiss years: Correspondence, 1902-1914}\ (\bibinfo  {publisher}
  {Princeton University Press},\ \bibinfo {year} {1993})\ p.~\bibinfo {pages}
  {44}\BibitemShut {NoStop}%
\bibitem [{\citenamefont {Hooke}(2019)}]{hooke2019posthumous}%
  \BibitemOpen
  \bibfield  {author} {\bibinfo {author} {\bibfnamefont {R.}~\bibnamefont
  {Hooke}},\ }\bibinfo {title} {The posthumous works of {R}obert {H}ooke}\
  (\bibinfo  {publisher} {Routledge},\ \bibinfo {year} {2019})\ p.\ \bibinfo
  {pages} {116}\BibitemShut {NoStop}%
\bibitem [{\citenamefont {Kubo}(1966)}]{Kubo_1966}%
  \BibitemOpen
  \bibfield  {author} {\bibinfo {author} {\bibfnamefont {R.}~\bibnamefont
  {Kubo}},\ }\bibfield  {title} {\bibinfo {title} {The fluctuation-dissipation
  theorem},\ }\href {https://doi.org/10.1088/0034-4885/29/1/306} {\bibfield
  {journal} {\bibinfo  {journal} {Rep. Prog. Phys.}\ }\textbf {\bibinfo
  {volume} {29}},\ \bibinfo {pages} {255} (\bibinfo {year} {1966})}\BibitemShut
  {NoStop}%
\bibitem [{\citenamefont {Lebowitz}\ and\ \citenamefont
  {Spohn}(1999)}]{lebowitz1999}%
  \BibitemOpen
  \bibfield  {author} {\bibinfo {author} {\bibfnamefont {J.~L.}\ \bibnamefont
  {Lebowitz}}\ and\ \bibinfo {author} {\bibfnamefont {H.}~\bibnamefont
  {Spohn}},\ }\bibfield  {title} {\bibinfo {title} {A gallavotti--cohen-type
  symmetry in the large deviation functional for stochastic dynamics},\
  }\href@noop {} {\bibfield  {journal} {\bibinfo  {journal} {J. Stat. Phys.}\
  }\textbf {\bibinfo {volume} {95}},\ \bibinfo {pages} {333} (\bibinfo {year}
  {1999})}\BibitemShut {NoStop}%
\bibitem [{\citenamefont {Maes}\ and\ \citenamefont
  {Neto{\v{c}}n{\`y}}(2003)}]{maes2003time}%
  \BibitemOpen
  \bibfield  {author} {\bibinfo {author} {\bibfnamefont {C.}~\bibnamefont
  {Maes}}\ and\ \bibinfo {author} {\bibfnamefont {K.}~\bibnamefont
  {Neto{\v{c}}n{\`y}}},\ }\bibfield  {title} {\bibinfo {title} {Time-reversal
  and entropy},\ }\href@noop {} {\bibfield  {journal} {\bibinfo  {journal} {J.
  Stat. Phys.}\ }\textbf {\bibinfo {volume} {110}},\ \bibinfo {pages} {269}
  (\bibinfo {year} {2003})}\BibitemShut {NoStop}%
\bibitem [{\citenamefont {Seifert}(2005)}]{Seifert2005}%
  \BibitemOpen
  \bibfield  {author} {\bibinfo {author} {\bibfnamefont {U.}~\bibnamefont
  {Seifert}},\ }\bibfield  {title} {\bibinfo {title} {Entropy production along
  a stochastic trajectory and an integral fluctuation theorem},\ }\href
  {https://link.aps.org/doi/10.1103/PhysRevLett.95.040602} {\bibfield
  {journal} {\bibinfo  {journal} {Phys. Rev. Lett.}\ }\textbf {\bibinfo
  {volume} {95}},\ \bibinfo {pages} {040602} (\bibinfo {year}
  {2005})}\BibitemShut {NoStop}%
\bibitem [{\citenamefont {Rold{\'{a}}n}\ and\ \citenamefont
  {Parrondo}(2012)}]{Rold_n_2012}%
  \BibitemOpen
  \bibfield  {author} {\bibinfo {author} {\bibfnamefont {{\'{E}
  }.}~\bibnamefont {Rold{\'{a}}n}}\ and\ \bibinfo {author} {\bibfnamefont
  {J.~M.~R.}\ \bibnamefont {Parrondo}},\ }\bibfield  {title} {\bibinfo {title}
  {Entropy production and {K}ullback-{L}eibler divergence between stationary
  trajectories of discrete systems},\ }\href
  {https://journals.aps.org/pre/abstract/10.1103/PhysRevE.85.031129} {\bibfield
   {journal} {\bibinfo  {journal} {Phys. Rev. E}\ }\textbf {\bibinfo {volume}
  {85}},\ \bibinfo {pages} {031129} (\bibinfo {year} {2012})}\BibitemShut
  {NoStop}%
\bibitem [{\citenamefont {Li}\ and\ \citenamefont {Cates}(2021)}]{Li_2021}%
  \BibitemOpen
  \bibfield  {author} {\bibinfo {author} {\bibfnamefont {Y.~I.}\ \bibnamefont
  {Li}}\ and\ \bibinfo {author} {\bibfnamefont {M.~E.}\ \bibnamefont {Cates}},\
  }\bibfield  {title} {\bibinfo {title} {Steady state entropy production rate
  for scalar {L}angevin field theories},\ }\href
  {https://doi.org/10.1088/1742-5468/abd311} {\bibfield  {journal} {\bibinfo
  {journal} {J. Stat. Mech. Theory Exp.}\ }\textbf {\bibinfo {volume} {2021}},\
  \bibinfo {pages} {013211} (\bibinfo {year} {2021})}\BibitemShut {NoStop}%
\bibitem [{\citenamefont {Nardini}\ \emph {et~al.}(2017)\citenamefont
  {Nardini}, \citenamefont {Fodor}, \citenamefont {Tjhung}, \citenamefont {van
  Wijland}, \citenamefont {Tailleur},\ and\ \citenamefont
  {Cates}}]{Nardini2017}%
  \BibitemOpen
  \bibfield  {author} {\bibinfo {author} {\bibfnamefont {C.}~\bibnamefont
  {Nardini}}, \bibinfo {author} {\bibfnamefont {{\'E}.}~\bibnamefont {Fodor}},
  \bibinfo {author} {\bibfnamefont {E.}~\bibnamefont {Tjhung}}, \bibinfo
  {author} {\bibfnamefont {F.}~\bibnamefont {van Wijland}}, \bibinfo {author}
  {\bibfnamefont {J.}~\bibnamefont {Tailleur}},\ and\ \bibinfo {author}
  {\bibfnamefont {M.~E.}\ \bibnamefont {Cates}},\ }\bibfield  {title} {\bibinfo
  {title} {Entropy production in field theories without time-reversal symmetry:
  Quantifying the non-equilibrium character of active matter},\ }\href
  {https://link.aps.org/doi/10.1103/PhysRevX.7.021007} {\bibfield  {journal}
  {\bibinfo  {journal} {Phys. Rev. X}\ }\textbf {\bibinfo {volume} {7}},\
  \bibinfo {pages} {021007} (\bibinfo {year} {2017})}\BibitemShut {NoStop}%
\bibitem [{\citenamefont {Seara}\ \emph {et~al.}(2021)\citenamefont {Seara},
  \citenamefont {Machta},\ and\ \citenamefont {Murrell}}]{seara2021}%
  \BibitemOpen
  \bibfield  {author} {\bibinfo {author} {\bibfnamefont {D.~S.}\ \bibnamefont
  {Seara}}, \bibinfo {author} {\bibfnamefont {B.~B.}\ \bibnamefont {Machta}},\
  and\ \bibinfo {author} {\bibfnamefont {M.~P.}\ \bibnamefont {Murrell}},\
  }\bibfield  {title} {\bibinfo {title} {Irreversibility in dynamical phases
  and transitions},\ }\href {https://doi.org/10.1038/s41467-020-20281-2}
  {\bibfield  {journal} {\bibinfo  {journal} {Nat. Commun.}\ }\textbf {\bibinfo
  {volume} {12}},\ \bibinfo {pages} {1} (\bibinfo {year} {2021})}\BibitemShut
  {NoStop}%
\bibitem [{\citenamefont {Borthne}\ \emph {et~al.}(2020)\citenamefont
  {Borthne}, \citenamefont {Fodor},\ and\ \citenamefont {Cates}}]{Borthne2020}%
  \BibitemOpen
  \bibfield  {author} {\bibinfo {author} {\bibfnamefont {{\O}.}~\bibnamefont
  {Borthne}}, \bibinfo {author} {\bibfnamefont {{\'E}.}~\bibnamefont {Fodor}},\
  and\ \bibinfo {author} {\bibfnamefont {M.}~\bibnamefont {Cates}},\ }\bibfield
   {title} {\bibinfo {title} {Time-reversal symmetry violations and entropy
  production in field theories of polar active matter},\ }\href
  {https://iopscience.iop.org/article/10.1088/1367-2630/abcd66} {\bibfield
  {journal} {\bibinfo  {journal} {New J. Phys.}\ }\textbf {\bibinfo {volume}
  {22}},\ \bibinfo {pages} {123012} (\bibinfo {year} {2020})}\BibitemShut
  {NoStop}%
\bibitem [{\citenamefont {Pruessner}\ and\ \citenamefont
  {Garcia-Millan}(2022)}]{pruessner2022field}%
  \BibitemOpen
  \bibfield  {author} {\bibinfo {author} {\bibfnamefont {G.}~\bibnamefont
  {Pruessner}}\ and\ \bibinfo {author} {\bibfnamefont {R.}~\bibnamefont
  {Garcia-Millan}},\ }\href@noop {} {\bibinfo {title} {Field theories of active
  particle systems and their entropy production}} (\bibinfo {year} {2022}),\
  \Eprint {https://arxiv.org/abs/2211.11906} {arXiv:2211.11906
  [cond-mat.stat-mech]} \BibitemShut {NoStop}%
\bibitem [{\citenamefont {Ro}\ \emph {et~al.}(2022)\citenamefont {Ro},
  \citenamefont {Guo}, \citenamefont {Shih}, \citenamefont {Phan},
  \citenamefont {Austin}, \citenamefont {Levine}, \citenamefont {Chaikin},\
  and\ \citenamefont {Martiniani}}]{ro2022model}%
  \BibitemOpen
  \bibfield  {author} {\bibinfo {author} {\bibfnamefont {S.}~\bibnamefont
  {Ro}}, \bibinfo {author} {\bibfnamefont {B.}~\bibnamefont {Guo}}, \bibinfo
  {author} {\bibfnamefont {A.}~\bibnamefont {Shih}}, \bibinfo {author}
  {\bibfnamefont {T.~V.}\ \bibnamefont {Phan}}, \bibinfo {author}
  {\bibfnamefont {R.~H.}\ \bibnamefont {Austin}}, \bibinfo {author}
  {\bibfnamefont {D.}~\bibnamefont {Levine}}, \bibinfo {author} {\bibfnamefont
  {P.~M.}\ \bibnamefont {Chaikin}},\ and\ \bibinfo {author} {\bibfnamefont
  {S.}~\bibnamefont {Martiniani}},\ }\bibfield  {title} {\bibinfo {title}
  {Model-free measurement of local entropy production and extractable work in
  active matter},\ }\href@noop {} {\bibfield  {journal} {\bibinfo  {journal}
  {Phys. Rev. Lett.}\ }\textbf {\bibinfo {volume} {129}},\ \bibinfo {pages}
  {220601} (\bibinfo {year} {2022})}\BibitemShut {NoStop}%
\bibitem [{\citenamefont {Tan}\ \emph {et~al.}(2021)\citenamefont {Tan},
  \citenamefont {Watson}, \citenamefont {Chao}, \citenamefont {Li},
  \citenamefont {Gingrich}, \citenamefont {Horowitz},\ and\ \citenamefont
  {Fakhri}}]{tan2021scale}%
  \BibitemOpen
  \bibfield  {author} {\bibinfo {author} {\bibfnamefont {T.~H.}\ \bibnamefont
  {Tan}}, \bibinfo {author} {\bibfnamefont {G.~A.}\ \bibnamefont {Watson}},
  \bibinfo {author} {\bibfnamefont {Y.-C.}\ \bibnamefont {Chao}}, \bibinfo
  {author} {\bibfnamefont {J.}~\bibnamefont {Li}}, \bibinfo {author}
  {\bibfnamefont {T.~R.}\ \bibnamefont {Gingrich}}, \bibinfo {author}
  {\bibfnamefont {J.~M.}\ \bibnamefont {Horowitz}},\ and\ \bibinfo {author}
  {\bibfnamefont {N.}~\bibnamefont {Fakhri}},\ }\href@noop {} {\bibinfo {title}
  {Scale-dependent irreversibility in living matter}} (\bibinfo {year}
  {2021}),\ \Eprint {https://arxiv.org/abs/2107.05701} {arXiv:2107.05701
  [physics.bio-ph]} \BibitemShut {NoStop}%
\bibitem [{\citenamefont {Battle}\ \emph {et~al.}(2016)\citenamefont {Battle},
  \citenamefont {Broedersz}, \citenamefont {Fakhri}, \citenamefont {Geyer},
  \citenamefont {Howard}, \citenamefont {Schmidt},\ and\ \citenamefont
  {MacKintosh}}]{battle2016broken}%
  \BibitemOpen
  \bibfield  {author} {\bibinfo {author} {\bibfnamefont {C.}~\bibnamefont
  {Battle}}, \bibinfo {author} {\bibfnamefont {C.~P.}\ \bibnamefont
  {Broedersz}}, \bibinfo {author} {\bibfnamefont {N.}~\bibnamefont {Fakhri}},
  \bibinfo {author} {\bibfnamefont {V.~F.}\ \bibnamefont {Geyer}}, \bibinfo
  {author} {\bibfnamefont {J.}~\bibnamefont {Howard}}, \bibinfo {author}
  {\bibfnamefont {C.~F.}\ \bibnamefont {Schmidt}},\ and\ \bibinfo {author}
  {\bibfnamefont {F.~C.}\ \bibnamefont {MacKintosh}},\ }\bibfield  {title}
  {\bibinfo {title} {Broken detailed balance at mesoscopic scales in active
  biological systems},\ }\href@noop {} {\bibfield  {journal} {\bibinfo
  {journal} {Science}\ }\textbf {\bibinfo {volume} {352}},\ \bibinfo {pages}
  {604} (\bibinfo {year} {2016})}\BibitemShut {NoStop}%
\bibitem [{\citenamefont {Bestehorn}\ \emph
  {et~al.}(1989{\natexlab{a}})\citenamefont {Bestehorn}, \citenamefont
  {Friedrich},\ and\ \citenamefont {Haken}}]{BESTEHORN1989}%
  \BibitemOpen
  \bibfield  {author} {\bibinfo {author} {\bibfnamefont {M.}~\bibnamefont
  {Bestehorn}}, \bibinfo {author} {\bibfnamefont {R.}~\bibnamefont
  {Friedrich}},\ and\ \bibinfo {author} {\bibfnamefont {H.}~\bibnamefont
  {Haken}},\ }\bibfield  {title} {\bibinfo {title} {Traveling waves in
  nonequilibrium systems},\ }\href
  {https://www.sciencedirect.com/science/article/pii/0167278989901371}
  {\bibfield  {journal} {\bibinfo  {journal} {Phys. D: Nonlinear Phenom.}\
  }\textbf {\bibinfo {volume} {37}},\ \bibinfo {pages} {295} (\bibinfo {year}
  {1989}{\natexlab{a}})}\BibitemShut {NoStop}%
\bibitem [{\citenamefont {Bestehorn}\ \emph
  {et~al.}(1989{\natexlab{b}})\citenamefont {Bestehorn}, \citenamefont
  {Friedrich},\ and\ \citenamefont {Haken}}]{bestehorn19892}%
  \BibitemOpen
  \bibfield  {author} {\bibinfo {author} {\bibfnamefont {M.}~\bibnamefont
  {Bestehorn}}, \bibinfo {author} {\bibfnamefont {R.}~\bibnamefont
  {Friedrich}},\ and\ \bibinfo {author} {\bibfnamefont {H.}~\bibnamefont
  {Haken}},\ }\bibfield  {title} {\bibinfo {title} {Two-dimensional traveling
  wave patterns in nonequilibrium systems},\ }\href@noop {} {\bibfield
  {journal} {\bibinfo  {journal} {Z. Phys. B.}\ }\textbf {\bibinfo {volume}
  {75}},\ \bibinfo {pages} {265} (\bibinfo {year}
  {1989}{\natexlab{b}})}\BibitemShut {NoStop}%
\bibitem [{\citenamefont {Ghosh}\ \emph {et~al.}(2021)\citenamefont {Ghosh},
  \citenamefont {Gutti},\ and\ \citenamefont {Chaudhuri}}]{Ghosh2021}%
  \BibitemOpen
  \bibfield  {author} {\bibinfo {author} {\bibfnamefont {S.}~\bibnamefont
  {Ghosh}}, \bibinfo {author} {\bibfnamefont {S.}~\bibnamefont {Gutti}},\ and\
  \bibinfo {author} {\bibfnamefont {D.}~\bibnamefont {Chaudhuri}},\ }\bibfield
  {title} {\bibinfo {title} {Pattern formation{,} localized and running
  pulsation on active spherical membranes},\ }\href
  {https://doi.org/10.1039/D1SM00937K} {\bibfield  {journal} {\bibinfo
  {journal} {Soft Matter}\ }\textbf {\bibinfo {volume} {17}},\ \bibinfo {pages}
  {10614} (\bibinfo {year} {2021})}\BibitemShut {NoStop}%
\bibitem [{\citenamefont {Ramaswamy}\ \emph {et~al.}(2000)\citenamefont
  {Ramaswamy}, \citenamefont {Toner},\ and\ \citenamefont
  {Prost}}]{Ramaswamy2000}%
  \BibitemOpen
  \bibfield  {author} {\bibinfo {author} {\bibfnamefont {S.}~\bibnamefont
  {Ramaswamy}}, \bibinfo {author} {\bibfnamefont {J.}~\bibnamefont {Toner}},\
  and\ \bibinfo {author} {\bibfnamefont {J.}~\bibnamefont {Prost}},\ }\bibfield
   {title} {\bibinfo {title} {Nonequilibrium fluctuations, traveling waves, and
  instabilities in active membranes},\ }\href
  {https://link.aps.org/doi/10.1103/PhysRevLett.84.3494} {\bibfield  {journal}
  {\bibinfo  {journal} {Phys. Rev. Lett.}\ }\textbf {\bibinfo {volume} {84}},\
  \bibinfo {pages} {3494} (\bibinfo {year} {2000})}\BibitemShut {NoStop}%
\bibitem [{\citenamefont {Agrawal}\ \emph {et~al.}(2017)\citenamefont
  {Agrawal}, \citenamefont {Bruss},\ and\ \citenamefont {Glotzer}}]{Mayank17}%
  \BibitemOpen
  \bibfield  {author} {\bibinfo {author} {\bibfnamefont {M.}~\bibnamefont
  {Agrawal}}, \bibinfo {author} {\bibfnamefont {I.~R.}\ \bibnamefont {Bruss}},\
  and\ \bibinfo {author} {\bibfnamefont {S.~C.}\ \bibnamefont {Glotzer}},\
  }\bibfield  {title} {\bibinfo {title} {Tunable emergent structures and
  traveling waves in mixtures of passive and contact-triggered-active
  particles},\ }\href {https://doi.org/10.1039/C7SM00888K} {\bibfield
  {journal} {\bibinfo  {journal} {Soft Matter}\ }\textbf {\bibinfo {volume}
  {13}},\ \bibinfo {pages} {6332} (\bibinfo {year} {2017})}\BibitemShut
  {NoStop}%
\bibitem [{\citenamefont {Frey}\ \emph {et~al.}(2018)\citenamefont {Frey},
  \citenamefont {Halatek}, \citenamefont {Kretschmer},\ and\ \citenamefont
  {Schwille}}]{frey2018protein}%
  \BibitemOpen
  \bibfield  {author} {\bibinfo {author} {\bibfnamefont {E.}~\bibnamefont
  {Frey}}, \bibinfo {author} {\bibfnamefont {J.}~\bibnamefont {Halatek}},
  \bibinfo {author} {\bibfnamefont {S.}~\bibnamefont {Kretschmer}},\ and\
  \bibinfo {author} {\bibfnamefont {P.}~\bibnamefont {Schwille}},\ }\bibfield
  {title} {\bibinfo {title} {Protein pattern formation},\ }in\ \href@noop {}
  {\emph {\bibinfo {booktitle} {Physics of biological membranes}}}\ (\bibinfo
  {publisher} {Springer},\ \bibinfo {year} {2018})\ pp.\ \bibinfo {pages}
  {229--260}\BibitemShut {NoStop}%
\bibitem [{\citenamefont {Huang}\ \emph {et~al.}(2003)\citenamefont {Huang},
  \citenamefont {Meir},\ and\ \citenamefont {Wingreen}}]{huang2003dynamic}%
  \BibitemOpen
  \bibfield  {author} {\bibinfo {author} {\bibfnamefont {K.~C.}\ \bibnamefont
  {Huang}}, \bibinfo {author} {\bibfnamefont {Y.}~\bibnamefont {Meir}},\ and\
  \bibinfo {author} {\bibfnamefont {N.~S.}\ \bibnamefont {Wingreen}},\
  }\bibfield  {title} {\bibinfo {title} {Dynamic structures in {E}scherichia
  coli: spontaneous formation of {M}in{E} rings and {M}in{D} polar zones},\
  }\href {https://www.pnas.org/doi/10.1073/pnas.2135445100} {\bibfield
  {journal} {\bibinfo  {journal} {Proc. Natl. Acad. Sci. U. S. A.}\ }\textbf
  {\bibinfo {volume} {100}},\ \bibinfo {pages} {12724} (\bibinfo {year}
  {2003})}\BibitemShut {NoStop}%
\bibitem [{\citenamefont {Amari}(1977)}]{Amari1977}%
  \BibitemOpen
  \bibfield  {author} {\bibinfo {author} {\bibfnamefont {S.-i.}\ \bibnamefont
  {Amari}},\ }\bibfield  {title} {\bibinfo {title} {Dynamic of pattern
  formation in lateral-inhibition type neural fields},\ }\href@noop {}
  {\bibfield  {journal} {\bibinfo  {journal} {Biol. Cybern.}\ }\textbf
  {\bibinfo {volume} {27}},\ \bibinfo {pages} {77} (\bibinfo {year}
  {1977})}\BibitemShut {NoStop}%
\bibitem [{\citenamefont {Wang}(2015)}]{Wang15}%
  \BibitemOpen
  \bibfield  {author} {\bibinfo {author} {\bibfnamefont {J.}~\bibnamefont
  {Wang}},\ }\bibfield  {title} {\bibinfo {title} {Landscape and flux theory of
  non-equilibrium dynamical systems with application to biology},\ }\href
  {https://www.tandfonline.com/doi/full/10.1080/00018732.2015.1037068}
  {\bibfield  {journal} {\bibinfo  {journal} {Adv. Phys.}\ }\textbf {\bibinfo
  {volume} {64}},\ \bibinfo {pages} {1} (\bibinfo {year} {2015})}\BibitemShut
  {NoStop}%
\bibitem [{\citenamefont {Bhattacharya}\ \emph {et~al.}(2020)\citenamefont
  {Bhattacharya}, \citenamefont {Banerjee}, \citenamefont {Miao}, \citenamefont
  {Zhan}, \citenamefont {Devreotes},\ and\ \citenamefont
  {Iglesias}}]{Bhattacharya20}%
  \BibitemOpen
  \bibfield  {author} {\bibinfo {author} {\bibfnamefont {S.}~\bibnamefont
  {Bhattacharya}}, \bibinfo {author} {\bibfnamefont {T.}~\bibnamefont
  {Banerjee}}, \bibinfo {author} {\bibfnamefont {Y.}~\bibnamefont {Miao}},
  \bibinfo {author} {\bibfnamefont {H.}~\bibnamefont {Zhan}}, \bibinfo {author}
  {\bibfnamefont {P.}~\bibnamefont {Devreotes}},\ and\ \bibinfo {author}
  {\bibfnamefont {P.}~\bibnamefont {Iglesias}},\ }\bibfield  {title} {\bibinfo
  {title} {Traveling and standing waves mediate pattern formation in cellular
  protrusions},\ }\href {https://www.science.org/doi/10.1126/sciadv.aay7682}
  {\bibfield  {journal} {\bibinfo  {journal} {Sci. Adv.}\ }\textbf {\bibinfo
  {volume} {6}},\ \bibinfo {pages} {eaay7682} (\bibinfo {year}
  {2020})}\BibitemShut {NoStop}%
\bibitem [{\citenamefont {Welch}\ and\ \citenamefont
  {Kaiser}(2002)}]{Welch2002}%
  \BibitemOpen
  \bibfield  {author} {\bibinfo {author} {\bibfnamefont {R.}~\bibnamefont
  {Welch}}\ and\ \bibinfo {author} {\bibfnamefont {D.}~\bibnamefont {Kaiser}},\
  }\bibfield  {title} {\bibinfo {title} {Cell behavior in traveling wave
  patterns of myxobacteria},\ }\href
  {https://www.pnas.org/doi/10.1073/pnas.261574598} {\bibfield  {journal}
  {\bibinfo  {journal} {Proc. Natl. Acad. Sci. U. S. A.}\ }\textbf {\bibinfo
  {volume} {98}},\ \bibinfo {pages} {14907} (\bibinfo {year}
  {2002})}\BibitemShut {NoStop}%
\bibitem [{\citenamefont {Takada}\ \emph {et~al.}(2022)\citenamefont {Takada},
  \citenamefont {Yoshinaga}, \citenamefont {Doi},\ and\ \citenamefont
  {Fujiwara}}]{Takada22}%
  \BibitemOpen
  \bibfield  {author} {\bibinfo {author} {\bibfnamefont {S.}~\bibnamefont
  {Takada}}, \bibinfo {author} {\bibfnamefont {N.}~\bibnamefont {Yoshinaga}},
  \bibinfo {author} {\bibfnamefont {N.}~\bibnamefont {Doi}},\ and\ \bibinfo
  {author} {\bibfnamefont {K.}~\bibnamefont {Fujiwara}},\ }\bibfield  {title}
  {\bibinfo {title} {Mode selection mechanism in traveling and standing waves
  revealed by min wave reconstituted in artificial cells},\ }\href@noop {}
  {\bibfield  {journal} {\bibinfo  {journal} {Sci. Adv.}\ }\textbf {\bibinfo
  {volume} {8}},\ \bibinfo {pages} {eabm8460} (\bibinfo {year}
  {2022})}\BibitemShut {NoStop}%
\bibitem [{\citenamefont {Demarchi}\ \emph {et~al.}(2023)\citenamefont
  {Demarchi}, \citenamefont {Goychuk}, \citenamefont {Maryshev},\ and\
  \citenamefont {Frey}}]{demarchi2023enzyme}%
  \BibitemOpen
  \bibfield  {author} {\bibinfo {author} {\bibfnamefont {L.}~\bibnamefont
  {Demarchi}}, \bibinfo {author} {\bibfnamefont {A.}~\bibnamefont {Goychuk}},
  \bibinfo {author} {\bibfnamefont {I.}~\bibnamefont {Maryshev}},\ and\
  \bibinfo {author} {\bibfnamefont {E.}~\bibnamefont {Frey}},\ }\href@noop {}
  {\bibinfo {title} {Enzyme-enriched condensates show self-propulsion,
  positioning, and coexistence}} (\bibinfo {year} {2023}),\ \Eprint
  {https://arxiv.org/abs/2301.00392} {arXiv:2301.00392 [physics.bio-ph]}
  \BibitemShut {NoStop}%
\bibitem [{\citenamefont {Goldbeter}(2018)}]{Goldbeter18}%
  \BibitemOpen
  \bibfield  {author} {\bibinfo {author} {\bibfnamefont {A.}~\bibnamefont
  {Goldbeter}},\ }\bibfield  {title} {\bibinfo {title} {Dissipative structures
  in biological systems: bistability, oscillations, spatial patterns and
  waves},\ }\href@noop {} {\bibfield  {journal} {\bibinfo  {journal} {Philos.
  Trans. R. Soc. A}\ }\textbf {\bibinfo {volume} {376}},\ \bibinfo {pages}
  {20170376} (\bibinfo {year} {2018})}\BibitemShut {NoStop}%
\bibitem [{\citenamefont {Cross}\ and\ \citenamefont
  {Hohenberg}(1993)}]{Cross93}%
  \BibitemOpen
  \bibfield  {author} {\bibinfo {author} {\bibfnamefont {M.~C.}\ \bibnamefont
  {Cross}}\ and\ \bibinfo {author} {\bibfnamefont {P.~C.}\ \bibnamefont
  {Hohenberg}},\ }\bibfield  {title} {\bibinfo {title} {Pattern formation
  outside of equilibrium},\ }\href
  {https://link.aps.org/doi/10.1103/RevModPhys.65.851} {\bibfield  {journal}
  {\bibinfo  {journal} {Rev. Mod. Phys.}\ }\textbf {\bibinfo {volume} {65}},\
  \bibinfo {pages} {851} (\bibinfo {year} {1993})}\BibitemShut {NoStop}%
\bibitem [{\citenamefont {Tiezzi}\ \emph {et~al.}(2008)\citenamefont {Tiezzi},
  \citenamefont {Pulselli}, \citenamefont {Marchettini},\ and\ \citenamefont
  {Tiezzi}}]{Tiezzi2008}%
  \BibitemOpen
  \bibfield  {author} {\bibinfo {author} {\bibfnamefont {E.}~\bibnamefont
  {Tiezzi}}, \bibinfo {author} {\bibfnamefont {R.}~\bibnamefont {Pulselli}},
  \bibinfo {author} {\bibfnamefont {N.}~\bibnamefont {Marchettini}},\ and\
  \bibinfo {author} {\bibfnamefont {E.}~\bibnamefont {Tiezzi}},\ }\bibfield
  {title} {\bibinfo {title} {Dissipative structures in nature and human
  systems}\ }(\bibinfo {year} {2008})\ pp.\ \bibinfo {pages}
  {293--299}\BibitemShut {NoStop}%
\bibitem [{\citenamefont {Belintsev}(1983)}]{Belintsev_1983}%
  \BibitemOpen
  \bibfield  {author} {\bibinfo {author} {\bibfnamefont {B.~N.}\ \bibnamefont
  {Belintsev}},\ }\bibfield  {title} {\bibinfo {title} {Dissipative structures
  and the problem of biological pattern formation},\ }\href
  {https://doi.org/10.1070/pu1983v026n09abeh004492} {\bibfield  {journal}
  {\bibinfo  {journal} {Sov. phys., Usp.}\ }\textbf {\bibinfo {volume} {26}},\
  \bibinfo {pages} {775} (\bibinfo {year} {1983})}\BibitemShut {NoStop}%
\bibitem [{\citenamefont {Cross}\ and\ \citenamefont
  {Greenside}(2009)}]{cross_greenside_2009}%
  \BibitemOpen
  \bibfield  {author} {\bibinfo {author} {\bibfnamefont {M.}~\bibnamefont
  {Cross}}\ and\ \bibinfo {author} {\bibfnamefont {H.}~\bibnamefont
  {Greenside}},\ }\bibinfo {title} {Oscillatory patterns},\ in\ \href@noop {}
  {\emph {\bibinfo {booktitle} {Pattern Formation and Dynamics in
  Nonequilibrium Systems}}}\ (\bibinfo  {publisher} {Cambridge University
  Press},\ \bibinfo {year} {2009})\ p.\ \bibinfo {pages}
  {358–400}\BibitemShut {NoStop}%
\bibitem [{\citenamefont {Hanai}\ and\ \citenamefont
  {Littlewood}(2020)}]{hanai2020critical}%
  \BibitemOpen
  \bibfield  {author} {\bibinfo {author} {\bibfnamefont {R.}~\bibnamefont
  {Hanai}}\ and\ \bibinfo {author} {\bibfnamefont {P.~B.}\ \bibnamefont
  {Littlewood}},\ }\bibfield  {title} {\bibinfo {title} {Critical fluctuations
  at a many-body exceptional point},\ }\href
  {https://journals.aps.org/prresearch/abstract/10.1103/PhysRevResearch.2.033018}
  {\bibfield  {journal} {\bibinfo  {journal} {Phys. {R}ev. Research}\ }\textbf
  {\bibinfo {volume} {2}},\ \bibinfo {pages} {033018} (\bibinfo {year}
  {2020})}\BibitemShut {NoStop}%
\bibitem [{\citenamefont {Fruchart}\ \emph {et~al.}(2021)\citenamefont
  {Fruchart}, \citenamefont {Hanai}, \citenamefont {Littlewood},\ and\
  \citenamefont {Vitelli}}]{Fruchart2021}%
  \BibitemOpen
  \bibfield  {author} {\bibinfo {author} {\bibfnamefont {M.}~\bibnamefont
  {Fruchart}}, \bibinfo {author} {\bibfnamefont {R.}~\bibnamefont {Hanai}},
  \bibinfo {author} {\bibfnamefont {P.~B.}\ \bibnamefont {Littlewood}},\ and\
  \bibinfo {author} {\bibfnamefont {V.}~\bibnamefont {Vitelli}},\ }\bibfield
  {title} {\bibinfo {title} {Non-reciprocal phase transitions},\ }\href
  {http://dx.doi.org/10.1038/s41586-021-03375-9} {\bibfield  {journal}
  {\bibinfo  {journal} {Nature}\ }\textbf {\bibinfo {volume} {592}},\ \bibinfo
  {pages} {363–369} (\bibinfo {year} {2021})}\BibitemShut {NoStop}%
\bibitem [{\citenamefont {Suchanek}\ \emph
  {et~al.}(2023{\natexlab{a}})\citenamefont {Suchanek}, \citenamefont {Kroy},\
  and\ \citenamefont {Loos}}]{suchanek2023connection}%
  \BibitemOpen
  \bibfield  {author} {\bibinfo {author} {\bibfnamefont {T.}~\bibnamefont
  {Suchanek}}, \bibinfo {author} {\bibfnamefont {K.}~\bibnamefont {Kroy}},\
  and\ \bibinfo {author} {\bibfnamefont {S.~A.~M.}\ \bibnamefont {Loos}},\
  }\href@noop {} {\bibinfo {title} {Time-reversal and $\mathcal{PT}$ symmetry
  breaking in non-{H}ermitian field theories}} (\bibinfo {year}
  {2023}{\natexlab{a}}),\ \Eprint {https://arxiv.org/abs/arXiv:2305.05633}
  {arXiv:arXiv:2305.05633 [cond-mat.stat-mech]} \BibitemShut {NoStop}%
\bibitem [{\citenamefont {You}\ \emph {et~al.}(2020)\citenamefont {You},
  \citenamefont {Baskaran},\ and\ \citenamefont {Marchetti}}]{You_2020}%
  \BibitemOpen
  \bibfield  {author} {\bibinfo {author} {\bibfnamefont {Z.}~\bibnamefont
  {You}}, \bibinfo {author} {\bibfnamefont {A.}~\bibnamefont {Baskaran}},\ and\
  \bibinfo {author} {\bibfnamefont {M.~C.}\ \bibnamefont {Marchetti}},\
  }\bibfield  {title} {\bibinfo {title} {Nonreciprocity as a generic route to
  traveling states},\ }\href {http://dx.doi.org/10.1073/pnas.2010318117}
  {\bibfield  {journal} {\bibinfo  {journal} {Proc. Natl. Acad. Sci. U. S. A.}\
  }\textbf {\bibinfo {volume} {117}},\ \bibinfo {pages} {19767–19772}
  (\bibinfo {year} {2020})}\BibitemShut {NoStop}%
\bibitem [{\citenamefont {Saha}\ \emph {et~al.}(2020)\citenamefont {Saha},
  \citenamefont {Agudo-Canalejo},\ and\ \citenamefont
  {Golestanian}}]{Saha_2020}%
  \BibitemOpen
  \bibfield  {author} {\bibinfo {author} {\bibfnamefont {S.}~\bibnamefont
  {Saha}}, \bibinfo {author} {\bibfnamefont {J.}~\bibnamefont
  {Agudo-Canalejo}},\ and\ \bibinfo {author} {\bibfnamefont {R.}~\bibnamefont
  {Golestanian}},\ }\bibfield  {title} {\bibinfo {title} {Scalar active
  mixtures: The nonreciprocal {C}ahn-{H}illiard model},\ }\href
  {http://dx.doi.org/10.1103/PhysRevX.10.041009} {\bibfield  {journal}
  {\bibinfo  {journal} {Phys. Rev. X}\ }\textbf {\bibinfo {volume} {10}},\
  \bibinfo {pages} {041009} (\bibinfo {year} {2020})}\BibitemShut {NoStop}%
\bibitem [{\citenamefont {Frohoff-H\"ulsmann}\ \emph
  {et~al.}(2021)\citenamefont {Frohoff-H\"ulsmann}, \citenamefont {Wrembel},\
  and\ \citenamefont {Thiele}}]{frohoff2021suppression}%
  \BibitemOpen
  \bibfield  {author} {\bibinfo {author} {\bibfnamefont {T.}~\bibnamefont
  {Frohoff-H\"ulsmann}}, \bibinfo {author} {\bibfnamefont {J.}~\bibnamefont
  {Wrembel}},\ and\ \bibinfo {author} {\bibfnamefont {U.}~\bibnamefont
  {Thiele}},\ }\bibfield  {title} {\bibinfo {title} {Suppression of coarsening
  and emergence of oscillatory behavior in a {C}ahn-{H}illiard model with
  nonvariational coupling},\ }\href
  {https://link.aps.org/doi/10.1103/PhysRevE.103.042602} {\bibfield  {journal}
  {\bibinfo  {journal} {Phys. Rev. E}\ }\textbf {\bibinfo {volume} {103}},\
  \bibinfo {pages} {042602} (\bibinfo {year} {2021})}\BibitemShut {NoStop}%
\bibitem [{\citenamefont {Frohoff-H{\"u}lsmann}\ and\ \citenamefont
  {Thiele}(2023)}]{frohoff2023nonreciprocal}%
  \BibitemOpen
  \bibfield  {author} {\bibinfo {author} {\bibfnamefont {T.}~\bibnamefont
  {Frohoff-H{\"u}lsmann}}\ and\ \bibinfo {author} {\bibfnamefont
  {U.}~\bibnamefont {Thiele}},\ }\href@noop {} {\bibinfo {title} {Nonreciprocal
  {C}ahn-{H}illiard equations emerging as one of eight universal amplitude
  equations}} (\bibinfo {year} {2023}),\ \Eprint
  {https://arxiv.org/abs/2301.05568} {arXiv:2301.05568 [cond-mat.soft]}
  \BibitemShut {NoStop}%
\bibitem [{\citenamefont {Mandal}\ \emph {et~al.}(2022)\citenamefont {Mandal},
  \citenamefont {Jaramillo},\ and\ \citenamefont {Sollich}}]{Mandal22}%
  \BibitemOpen
  \bibfield  {author} {\bibinfo {author} {\bibfnamefont {R.}~\bibnamefont
  {Mandal}}, \bibinfo {author} {\bibfnamefont {S.~S.}\ \bibnamefont
  {Jaramillo}},\ and\ \bibinfo {author} {\bibfnamefont {P.}~\bibnamefont
  {Sollich}},\ }\href {https://arxiv.org/abs/2212.05618} {\bibinfo {title}
  {Robustness of travelling states in generic non-reciprocal mixtures}}
  (\bibinfo {year} {2022}),\ \Eprint {https://arxiv.org/abs/2212.05618}
  {arXiv:2212.05618 [cond-mat.stat-mech]} \BibitemShut {NoStop}%
\bibitem [{\citenamefont {Loos}\ and\ \citenamefont
  {Klapp}(2020)}]{loos2020irreversibility}%
  \BibitemOpen
  \bibfield  {author} {\bibinfo {author} {\bibfnamefont {S.~A.~M.}\
  \bibnamefont {Loos}}\ and\ \bibinfo {author} {\bibfnamefont {S.~H.~L.}\
  \bibnamefont {Klapp}},\ }\bibfield  {title} {\bibinfo {title}
  {Irreversibility, heat and information flows induced by non-reciprocal
  interactions},\ }\href
  {https://iopscience.iop.org/article/10.1088/1367-2630/abcc1e} {\bibfield
  {journal} {\bibinfo  {journal} {New J. Phys.}\ }\textbf {\bibinfo {volume}
  {22}},\ \bibinfo {pages} {123051} (\bibinfo {year} {2020})}\BibitemShut
  {NoStop}%
\bibitem [{\citenamefont {El-Ganainy}\ \emph {et~al.}(2018)\citenamefont
  {El-Ganainy}, \citenamefont {Makris}, \citenamefont {Khajavikhan},
  \citenamefont {Musslimani}, \citenamefont {Rotter},\ and\ \citenamefont
  {Christodoulides}}]{el2018non}%
  \BibitemOpen
  \bibfield  {author} {\bibinfo {author} {\bibfnamefont {R.}~\bibnamefont
  {El-Ganainy}}, \bibinfo {author} {\bibfnamefont {K.~G.}\ \bibnamefont
  {Makris}}, \bibinfo {author} {\bibfnamefont {M.}~\bibnamefont {Khajavikhan}},
  \bibinfo {author} {\bibfnamefont {Z.~H.}\ \bibnamefont {Musslimani}},
  \bibinfo {author} {\bibfnamefont {S.}~\bibnamefont {Rotter}},\ and\ \bibinfo
  {author} {\bibfnamefont {D.~N.}\ \bibnamefont {Christodoulides}},\ }\bibfield
   {title} {\bibinfo {title} {Non-hermitian physics and pt symmetry},\
  }\href@noop {} {\bibfield  {journal} {\bibinfo  {journal} {Nat. Phys.}\
  }\textbf {\bibinfo {volume} {14}},\ \bibinfo {pages} {11} (\bibinfo {year}
  {2018})}\BibitemShut {NoStop}%
\bibitem [{\citenamefont {Krasnok}\ \emph {et~al.}(2021)\citenamefont
  {Krasnok}, \citenamefont {Nefedkin},\ and\ \citenamefont
  {Alu}}]{krasnok2021paritytime}%
  \BibitemOpen
  \bibfield  {author} {\bibinfo {author} {\bibfnamefont {A.}~\bibnamefont
  {Krasnok}}, \bibinfo {author} {\bibfnamefont {N.}~\bibnamefont {Nefedkin}},\
  and\ \bibinfo {author} {\bibfnamefont {A.}~\bibnamefont {Alu}},\ }\href@noop
  {} {\bibinfo {title} {Parity-time symmetry and exceptional points: A
  tutorial}} (\bibinfo {year} {2021}),\ \Eprint
  {https://arxiv.org/abs/2103.08135} {arXiv:2103.08135} \BibitemShut {NoStop}%
\bibitem [{\citenamefont {Suchanek}\ \emph
  {et~al.}(2023{\natexlab{b}})\citenamefont {Suchanek}, \citenamefont {Kroy},\
  and\ \citenamefont {Loos}}]{suchanek2023entropy}%
  \BibitemOpen
  \bibfield  {author} {\bibinfo {author} {\bibfnamefont {T.}~\bibnamefont
  {Suchanek}}, \bibinfo {author} {\bibfnamefont {K.}~\bibnamefont {Kroy}},\
  and\ \bibinfo {author} {\bibfnamefont {S.~A.~M.}\ \bibnamefont {Loos}},\
  }\href@noop {} {\bibinfo {title} {Entropy production rate in the
  nonreciprocal {C}ahn-{H}illiard model}} (\bibinfo {year}
  {2023}{\natexlab{b}}),\ \Eprint {https://arxiv.org/abs/arXiv:2305.00744}
  {arXiv:arXiv:2305.00744 [cond-mat.soft]} \BibitemShut {NoStop}%
\bibitem [{\citenamefont {Kne{\v{z}}evi{\'c}}\ \emph
  {et~al.}(2022)\citenamefont {Kne{\v{z}}evi{\'c}}, \citenamefont {Welker},\
  and\ \citenamefont {Stark}}]{knevzevic2022}%
  \BibitemOpen
  \bibfield  {author} {\bibinfo {author} {\bibfnamefont {M.}~\bibnamefont
  {Kne{\v{z}}evi{\'c}}}, \bibinfo {author} {\bibfnamefont {T.}~\bibnamefont
  {Welker}},\ and\ \bibinfo {author} {\bibfnamefont {H.}~\bibnamefont
  {Stark}},\ }\bibfield  {title} {\bibinfo {title} {Collective motion of active
  particles exhibiting non-reciprocal orientational interactions},\ }\href
  {https://www.nature.com/articles/s41598-022-23597-9} {\bibfield  {journal}
  {\bibinfo  {journal} {Scientific Reports}\ }\textbf {\bibinfo {volume}
  {12}},\ \bibinfo {pages} {19437} (\bibinfo {year} {2022})}\BibitemShut
  {NoStop}%
\bibitem [{\citenamefont {Coullet}\ \emph {et~al.}(1989)\citenamefont
  {Coullet}, \citenamefont {Goldstein},\ and\ \citenamefont
  {Gunaratne}}]{Coullet89}%
  \BibitemOpen
  \bibfield  {author} {\bibinfo {author} {\bibfnamefont {P.}~\bibnamefont
  {Coullet}}, \bibinfo {author} {\bibfnamefont {R.~E.}\ \bibnamefont
  {Goldstein}},\ and\ \bibinfo {author} {\bibfnamefont {G.~H.}\ \bibnamefont
  {Gunaratne}},\ }\bibfield  {title} {\bibinfo {title} {Parity-breaking
  transitions of modulated patterns in hydrodynamic systems},\ }\href
  {https://link.aps.org/doi/10.1103/PhysRevLett.63.1954} {\bibfield  {journal}
  {\bibinfo  {journal} {Phys. Rev. Lett.}\ }\textbf {\bibinfo {volume} {63}},\
  \bibinfo {pages} {1954} (\bibinfo {year} {1989})}\BibitemShut {NoStop}%
\bibitem [{\citenamefont {Cummins}\ \emph {et~al.}(1993)\citenamefont
  {Cummins}, \citenamefont {Fourtune},\ and\ \citenamefont
  {Rabaud}}]{Cummins93}%
  \BibitemOpen
  \bibfield  {author} {\bibinfo {author} {\bibfnamefont {H.~Z.}\ \bibnamefont
  {Cummins}}, \bibinfo {author} {\bibfnamefont {L.}~\bibnamefont {Fourtune}},\
  and\ \bibinfo {author} {\bibfnamefont {M.}~\bibnamefont {Rabaud}},\
  }\bibfield  {title} {\bibinfo {title} {Successive bifurcations in directional
  viscous fingering},\ }\href
  {https://link.aps.org/doi/10.1103/PhysRevE.47.1727} {\bibfield  {journal}
  {\bibinfo  {journal} {Phys. Rev. E}\ }\textbf {\bibinfo {volume} {47}},\
  \bibinfo {pages} {1727} (\bibinfo {year} {1993})}\BibitemShut {NoStop}%
\bibitem [{\citenamefont {Pan}\ and\ \citenamefont {de~Bruyn}(1994)}]{Pan94}%
  \BibitemOpen
  \bibfield  {author} {\bibinfo {author} {\bibfnamefont {L.}~\bibnamefont
  {Pan}}\ and\ \bibinfo {author} {\bibfnamefont {J.~R.}\ \bibnamefont
  {de~Bruyn}},\ }\bibfield  {title} {\bibinfo {title} {Spatially uniform
  traveling cellular patterns at a driven interface},\ }\href
  {https://link.aps.org/doi/10.1103/PhysRevE.49.483} {\bibfield  {journal}
  {\bibinfo  {journal} {Phys. Rev. E}\ }\textbf {\bibinfo {volume} {49}},\
  \bibinfo {pages} {483} (\bibinfo {year} {1994})}\BibitemShut {NoStop}%
\bibitem [{\citenamefont {Cont}\ and\ \citenamefont {Fournie}(2010)}]{Itofunc}%
  \BibitemOpen
  \bibfield  {author} {\bibinfo {author} {\bibfnamefont {R.}~\bibnamefont
  {Cont}}\ and\ \bibinfo {author} {\bibfnamefont {D.}~\bibnamefont {Fournie}},\
  }\bibfield  {title} {\bibinfo {title} {A functional extension of the ito
  formula},\ }\href
  {https://www.sciencedirect.com/science/article/pii/S1631073X09003951}
  {\bibfield  {journal} {\bibinfo  {journal} {Comptes Rendus Math.}\ }\textbf
  {\bibinfo {volume} {348}},\ \bibinfo {pages} {57} (\bibinfo {year}
  {2010})}\BibitemShut {NoStop}%
\bibitem [{\citenamefont {Ashida}\ \emph {et~al.}(2020)\citenamefont {Ashida},
  \citenamefont {Gong},\ and\ \citenamefont {Ueda}}]{Ashida2020}%
  \BibitemOpen
  \bibfield  {author} {\bibinfo {author} {\bibfnamefont {Y.}~\bibnamefont
  {Ashida}}, \bibinfo {author} {\bibfnamefont {Z.}~\bibnamefont {Gong}},\ and\
  \bibinfo {author} {\bibfnamefont {M.}~\bibnamefont {Ueda}},\ }\bibfield
  {title} {\bibinfo {title} {Non-{H}ermitian physics},\ }\href
  {http://dx.doi.org/10.1080/00018732.2021.1876991} {\bibfield  {journal}
  {\bibinfo  {journal} {Adv. Phys.}\ }\textbf {\bibinfo {volume} {69}},\
  \bibinfo {pages} {249–435} (\bibinfo {year} {2020})}\BibitemShut {NoStop}%
\bibitem [{\citenamefont {Zelle}\ \emph {et~al.}(2023)\citenamefont {Zelle},
  \citenamefont {Daviet}, \citenamefont {Rosch},\ and\ \citenamefont
  {Diehl}}]{zelle2023universal}%
  \BibitemOpen
  \bibfield  {author} {\bibinfo {author} {\bibfnamefont {C.~P.}\ \bibnamefont
  {Zelle}}, \bibinfo {author} {\bibfnamefont {R.}~\bibnamefont {Daviet}},
  \bibinfo {author} {\bibfnamefont {A.}~\bibnamefont {Rosch}},\ and\ \bibinfo
  {author} {\bibfnamefont {S.}~\bibnamefont {Diehl}},\ }\href@noop {} {\bibinfo
  {title} {Universal phenomenology at critical exceptional points of
  nonequilibrium $o(n)$ models}} (\bibinfo {year} {2023}),\ \Eprint
  {https://arxiv.org/abs/2304.09207} {arXiv:2304.09207 [cond-mat.stat-mech]}
  \BibitemShut {NoStop}%
\bibitem [{Note1()}]{Note1}%
  \BibitemOpen
  \bibinfo {note} {See Ref.~\cite {suchanek2023connection} for the precise
  definition of the projector $\protect \hat {P}_0$}\BibitemShut {NoStop}%
\bibitem [{\citenamefont {Gardiner}(2009)}]{gardiner2009}%
  \BibitemOpen
  \bibfield  {author} {\bibinfo {author} {\bibfnamefont {C.}~\bibnamefont
  {Gardiner}},\ }\href@noop {} {\emph {\bibinfo {title} {Stochastic
  methods}}},\ Vol.~\bibinfo {volume} {4}\ (\bibinfo  {publisher} {Springer
  Berlin},\ \bibinfo {year} {2009})\BibitemShut {NoStop}%
\bibitem [{\citenamefont {Fodor}\ \emph {et~al.}(2016)\citenamefont {Fodor},
  \citenamefont {Nardini}, \citenamefont {Cates}, \citenamefont {Tailleur},
  \citenamefont {Visco},\ and\ \citenamefont {van Wijland}}]{fodor2016far}%
  \BibitemOpen
  \bibfield  {author} {\bibinfo {author} {\bibfnamefont {{\'E}.}~\bibnamefont
  {Fodor}}, \bibinfo {author} {\bibfnamefont {C.}~\bibnamefont {Nardini}},
  \bibinfo {author} {\bibfnamefont {M.~E.}\ \bibnamefont {Cates}}, \bibinfo
  {author} {\bibfnamefont {J.}~\bibnamefont {Tailleur}}, \bibinfo {author}
  {\bibfnamefont {P.}~\bibnamefont {Visco}},\ and\ \bibinfo {author}
  {\bibfnamefont {F.}~\bibnamefont {van Wijland}},\ }\bibfield  {title}
  {\bibinfo {title} {How far from equilibrium is active matter?},\ }\href@noop
  {} {\bibfield  {journal} {\bibinfo  {journal} {Physical {R}ev. {L}ett.}\
  }\textbf {\bibinfo {volume} {117}},\ \bibinfo {pages} {038103} (\bibinfo
  {year} {2016})}\BibitemShut {NoStop}%
\bibitem [{\citenamefont {Caprini}\ \emph {et~al.}(2019)\citenamefont
  {Caprini}, \citenamefont {Marconi}, \citenamefont {Puglisi},\ and\
  \citenamefont {Vulpiani}}]{Caprini_2019}%
  \BibitemOpen
  \bibfield  {author} {\bibinfo {author} {\bibfnamefont {L.}~\bibnamefont
  {Caprini}}, \bibinfo {author} {\bibfnamefont {U.~M.~B.}\ \bibnamefont
  {Marconi}}, \bibinfo {author} {\bibfnamefont {A.}~\bibnamefont {Puglisi}},\
  and\ \bibinfo {author} {\bibfnamefont {A.}~\bibnamefont {Vulpiani}},\
  }\bibfield  {title} {\bibinfo {title} {The entropy production of
  {O}rnstein{\textendash}{U}hlenbeck active particles: a path integral method
  for correlations},\ }\href@noop {} {\bibfield  {journal} {\bibinfo  {journal}
  {J. Stat. Mech. Theory Exp.}\ }\textbf {\bibinfo {volume} {2019}},\ \bibinfo
  {pages} {053203} (\bibinfo {year} {2019})}\BibitemShut {NoStop}%
\bibitem [{\citenamefont {Falasco}\ \emph {et~al.}(2016)\citenamefont
  {Falasco}, \citenamefont {Pfaller}, \citenamefont {Bregulla}, \citenamefont
  {Cichos},\ and\ \citenamefont {Kroy}}]{Falasco16}%
  \BibitemOpen
  \bibfield  {author} {\bibinfo {author} {\bibfnamefont {G.}~\bibnamefont
  {Falasco}}, \bibinfo {author} {\bibfnamefont {R.}~\bibnamefont {Pfaller}},
  \bibinfo {author} {\bibfnamefont {A.~P.}\ \bibnamefont {Bregulla}}, \bibinfo
  {author} {\bibfnamefont {F.}~\bibnamefont {Cichos}},\ and\ \bibinfo {author}
  {\bibfnamefont {K.}~\bibnamefont {Kroy}},\ }\bibfield  {title} {\bibinfo
  {title} {Exact symmetries in the velocity fluctuations of a hot {B}rownian
  swimmer},\ }\href {https://link.aps.org/doi/10.1103/PhysRevE.94.030602}
  {\bibfield  {journal} {\bibinfo  {journal} {Phys. Rev. E}\ }\textbf {\bibinfo
  {volume} {94}},\ \bibinfo {pages} {030602(R)} (\bibinfo {year}
  {2016})}\BibitemShut {NoStop}%
\bibitem [{Note2()}]{Note2}%
  \BibitemOpen
  \bibinfo {note} {The MSD of $\theta _c$ is readily obtained by applying the
  results from Eq.~\cite {Nguyen21} to Eq.~\protect \textup {\hbox
  {\mathsurround \z@ \protect \normalfont (\ignorespaces \ref
  {eq:centerofmass}\unskip \@@italiccorr )}}. We further note that the MSD of
  the mean phase $\theta _\protect \mathrm {m}\equiv (\theta _A^1+\protect
  \tilde \theta _B^{1})/2$, which may be more accessible from an experimental
  perspective, exhibits the same ballistic short-time regime and the identical,
  enhanced long-time diffusion coefficient.}\BibitemShut {Stop}%
\bibitem [{\citenamefont {Shankar}\ and\ \citenamefont
  {Marchetti}(2018)}]{shankar2018hidden}%
  \BibitemOpen
  \bibfield  {author} {\bibinfo {author} {\bibfnamefont {S.}~\bibnamefont
  {Shankar}}\ and\ \bibinfo {author} {\bibfnamefont {M.~C.}\ \bibnamefont
  {Marchetti}},\ }\bibfield  {title} {\bibinfo {title} {Hidden entropy
  production and work fluctuations in an ideal active gas},\ }\href@noop {}
  {\bibfield  {journal} {\bibinfo  {journal} {Phys. Rev. E}\ }\textbf {\bibinfo
  {volume} {98}},\ \bibinfo {pages} {020604(R)} (\bibinfo {year}
  {2018})}\BibitemShut {NoStop}%
\bibitem [{\citenamefont {Falasco}\ \emph {et~al.}(2018)\citenamefont
  {Falasco}, \citenamefont {Rao},\ and\ \citenamefont
  {Esposito}}]{falasco2018information}%
  \BibitemOpen
  \bibfield  {author} {\bibinfo {author} {\bibfnamefont {G.}~\bibnamefont
  {Falasco}}, \bibinfo {author} {\bibfnamefont {R.}~\bibnamefont {Rao}},\ and\
  \bibinfo {author} {\bibfnamefont {M.}~\bibnamefont {Esposito}},\ }\bibfield
  {title} {\bibinfo {title} {Information thermodynamics of {T}uring patterns},\
  }\href@noop {} {\bibfield  {journal} {\bibinfo  {journal} {Phys. Rev. Lett.}\
  }\textbf {\bibinfo {volume} {121}},\ \bibinfo {pages} {108301} (\bibinfo
  {year} {2018})}\BibitemShut {NoStop}%
\bibitem [{\citenamefont {Paoluzzi}(2022)}]{Paoluzzi22}%
  \BibitemOpen
  \bibfield  {author} {\bibinfo {author} {\bibfnamefont {M.}~\bibnamefont
  {Paoluzzi}},\ }\bibfield  {title} {\bibinfo {title} {Scaling of the entropy
  production rate in a ${\ensuremath{\varphi}}^{4}$ model of active matter},\
  }\href {https://doi.org/10.1103/PhysRevE.105.044139} {\bibfield  {journal}
  {\bibinfo  {journal} {Phys. Rev. E}\ }\textbf {\bibinfo {volume} {105}},\
  \bibinfo {pages} {044139} (\bibinfo {year} {2022})}\BibitemShut {NoStop}%
\bibitem [{\citenamefont {Caballero}\ and\ \citenamefont
  {Cates}(2020)}]{caballero2020stealth}%
  \BibitemOpen
  \bibfield  {author} {\bibinfo {author} {\bibfnamefont {F.}~\bibnamefont
  {Caballero}}\ and\ \bibinfo {author} {\bibfnamefont {M.~E.}\ \bibnamefont
  {Cates}},\ }\bibfield  {title} {\bibinfo {title} {Stealth entropy production
  in active field theories near {Ising} critical points},\ }\href
  {https://doi.org/10.1103/PhysRevLett.124.240604} {\bibfield  {journal}
  {\bibinfo  {journal} {Phys. Rev. Lett.}\ }\textbf {\bibinfo {volume} {124}},\
  \bibinfo {pages} {240604} (\bibinfo {year} {2020})}\BibitemShut {NoStop}%
\bibitem [{\citenamefont {Barato}\ and\ \citenamefont
  {Seifert}(2015)}]{Barato15}%
  \BibitemOpen
  \bibfield  {author} {\bibinfo {author} {\bibfnamefont {A.~C.}\ \bibnamefont
  {Barato}}\ and\ \bibinfo {author} {\bibfnamefont {U.}~\bibnamefont
  {Seifert}},\ }\bibfield  {title} {\bibinfo {title} {Thermodynamic uncertainty
  relation for biomolecular processes},\ }\href
  {https://doi.org/10.1103/PhysRevLett.114.158101} {\bibfield  {journal}
  {\bibinfo  {journal} {Phys. Rev. Lett.}\ }\textbf {\bibinfo {volume} {114}},\
  \bibinfo {pages} {158101} (\bibinfo {year} {2015})}\BibitemShut {NoStop}%
\bibitem [{\citenamefont {Gingrich}\ \emph {et~al.}(2016)\citenamefont
  {Gingrich}, \citenamefont {Horowitz}, \citenamefont {Perunov},\ and\
  \citenamefont {England}}]{Gingrich16}%
  \BibitemOpen
  \bibfield  {author} {\bibinfo {author} {\bibfnamefont {T.~R.}\ \bibnamefont
  {Gingrich}}, \bibinfo {author} {\bibfnamefont {J.~M.}\ \bibnamefont
  {Horowitz}}, \bibinfo {author} {\bibfnamefont {N.}~\bibnamefont {Perunov}},\
  and\ \bibinfo {author} {\bibfnamefont {J.~L.}\ \bibnamefont {England}},\
  }\bibfield  {title} {\bibinfo {title} {Dissipation bounds all steady-state
  current fluctuations},\ }\href
  {https://doi.org/10.1103/PhysRevLett.116.120601} {\bibfield  {journal}
  {\bibinfo  {journal} {Phys. Rev. Lett.}\ }\textbf {\bibinfo {volume} {116}},\
  \bibinfo {pages} {120601} (\bibinfo {year} {2016})}\BibitemShut {NoStop}%
\bibitem [{\citenamefont {Niggemann}\ and\ \citenamefont
  {Seifert}(2020)}]{niggemann2020field}%
  \BibitemOpen
  \bibfield  {author} {\bibinfo {author} {\bibfnamefont {O.}~\bibnamefont
  {Niggemann}}\ and\ \bibinfo {author} {\bibfnamefont {U.}~\bibnamefont
  {Seifert}},\ }\bibfield  {title} {\bibinfo {title} {Field-theoretic
  thermodynamic uncertainty relation: General formulation exemplified with the
  kardar--parisi--zhang equation},\ }\href@noop {} {\bibfield  {journal}
  {\bibinfo  {journal} {J. Stat. Phys.}\ }\textbf {\bibinfo {volume} {178}},\
  \bibinfo {pages} {1142} (\bibinfo {year} {2020})}\BibitemShut {NoStop}%
\bibitem [{\citenamefont {Alston}\ \emph {et~al.}(2023)\citenamefont {Alston},
  \citenamefont {Cocconi},\ and\ \citenamefont
  {Bertrand}}]{alston2023irreversibility}%
  \BibitemOpen
  \bibfield  {author} {\bibinfo {author} {\bibfnamefont {H.}~\bibnamefont
  {Alston}}, \bibinfo {author} {\bibfnamefont {L.}~\bibnamefont {Cocconi}},\
  and\ \bibinfo {author} {\bibfnamefont {T.}~\bibnamefont {Bertrand}},\
  }\href@noop {} {\bibinfo {title} {Irreversibility across a nonreciprocal
  $\mathcal{PT}$-symmetry-breaking phase transition}} (\bibinfo {year}
  {2023}),\ \Eprint {https://arxiv.org/abs/2304.08661} {arXiv:2304.08661
  [cond-mat.stat-mech]} \BibitemShut {NoStop}%
\end{thebibliography}%

\end{document}